\pdfoutput=1
\documentclass[aip,cha,reprint,amsmath,amssymb,floatfix]{revtex4-2}
\usepackage{graphicx}
\usepackage{hyperref}
\hypersetup{hidelinks}
\makeatletter
\let\sm@hangfrom@section\@hangfrom@section
\newcommand{\supplementarytitle}[2]{%
  \onecolumngrid
  \clearpage
  \begingroup
    \frontmatter@setup
    \frontmatter@title@above
    {\frontmatter@title@format #1\par}%
    \frontmatter@title@below
    {\frontmatter@authorformat #2\par}%
    \frontmatter@RRAPformat{}%
    \addvspace{\frontmatter@preabstractspace}%
    {\dimen@\baselineskip\advance\dimen@-\prevdepth
     \ifdim\dimen@>\z@\vskip\dimen@\fi}%
    \addvspace{\frontmatter@postabstractspace}%
    \frontmatter@finalspace
  \endgroup
  \twocolumngrid
  \suppressfloats[t]%
  \let\@hangfrom@section\sm@hangfrom@section
  \let\@sectioncntformat\relax
  \@removefromreset{equation}{section}%
  \setcounter{section}{0}\setcounter{equation}{0}%
  \setcounter{figure}{0}\setcounter{table}{0}%
  \renewcommand{\thesection}{S\arabic{section}}%
  \renewcommand{\theequation}{S\arabic{equation}}%
  \renewcommand{\thefigure}{S\arabic{figure}}%
  \renewcommand{\thetable}{S\arabic{table}}%
  \def\theHsection{S\arabic{section}}%
  \def\theHequation{S\arabic{equation}}%
  \def\theHfigure{S\arabic{figure}}%
  \def\theHtable{S\arabic{table}}%
}
\newenvironment{supplementaryreferences}{%
  \let\sm@label\label
  \let\sm@ref\ref
  \renewcommand{\label}[1]{\sm@label{sm-##1}}%
  \renewcommand{\ref}[1]{\sm@ref{sm-##1}}%
}{}
\makeatother

\begin{document}

\title{Energy-selective control of noise-assisted multipulsing by weak optical seeding in the dissipative-soliton-resonance regime}

\author{Vladimir L. Kalashnikov}
\email{vladimir.kalashnikov@ntnu.no}
\affiliation{Department of Physics, Norwegian University of Science and Technology, 7491 Trondheim, Norway}

\author{Alexander Rudenkov}
\affiliation{Department of Physics, Norwegian University of Science and Technology, 7491 Trondheim, Norway}

\author{Evgeni Sorokin}
\affiliation{Institut f\"ur Photonik, TU Wien, Gu\ss hausstra\ss e 27/387, A-1040 Vienna, Austria}
\affiliation{ATLA lasers AS, Richard Birkelands vei 2B, 7034 Trondheim, Norway}

\author{Irina T. Sorokina}
\affiliation{Department of Physics, Norwegian University of Science and Technology, 7491 Trondheim, Norway}
\affiliation{ATLA lasers AS, Richard Birkelands vei 2B, 7034 Trondheim, Norway}

\begin{abstract}
We investigate how weak optical injection changes single-pulse operation in a laser in the dissipative-soliton-resonance regime. The energy scans show that a small injected field suppresses multipulsing and extends predominantly single-pulse operation to higher energies. Further increasing the seed power makes this extended range nonuniform: regions of enhanced multipulse formation alternate with regions of suppression. We analyze timing and optical-phase fluctuations of the injection field. At the lowest energies, these fluctuations cause continuous-wave and multipulse generation to dominate the laser output. At the highest energies, injection-field phase fluctuations promote multiple pulses, whereas time fluctuations suppress them.  Deterministic calculations without noise or injection show that both single pulses and prepared pulse pairs can persist under the same settings. We show that the spectral entropy of a dissipative soliton reflects changes in individual pulse spectra and pulse interference. It characterizes the probabilistic transition between single- and multipulse regimes, but it does not, by itself, predict in detail which regime the laser will form. Together, the results connect the energy-selective action of optical injection to pulse formation and structure, while distinguishing persistence of a prepared pulse from its formation out of a background.

\end{abstract}

\maketitle

\begin{quotation}
\bfseries
A laser can store more energy in a single dissipative soliton by stretching the pulse in the dissipative soliton resonance regime, but fluctuations can cause it to form several pulses instead. Without optical injection, this limits the energy range in which a single pulse is usually obtained. A small injected field broadens that range. Related effects of weak injection are known from semiconductor and fiber lasers, but they operate outside the dissipative soliton resonance. Increasing the seed power further has a different effect: it introduces energy intervals where multiple pulses are favored, separated by intervals where they are suppressed. This nonuniform response is the central result of our study. Calculations of prepared single pulses and pulse pairs show that both can persist under the same settings. Their spectra then reveal how the laser shares energy between soliton constituents and how interference records their arrangement. Optical injection thus controls which pulse configuration forms and, at higher seed power, the pulses' structure.

\end{quotation}

\section{Introduction}
\label{sec:introduction}

Dissipative solitons (DSs) are localized states maintained by balances of dispersion and nonlinearity and of gain and loss. They underlie energy-scalable ultrashort-pulse generation in mode-locked lasers and exemplify self-organization in driven systems \cite{grelu2012dissipative}. In normal group-delay dispersion (NGD), a strongly chirped pulse can store increasing energy by broadening while its peak power remains nearly clamped. This is the dissipative-soliton-resonance (DSR) scaling mechanism \cite{chang2008dsr,grelu2010dissipative}. Experiments nevertheless show transitions to multiple pulses \cite{wu2009dsr,sanchez2023overview}, motivating the distinction between the energy that one pulse can sustain and the pulse configuration that the laser actually reaches.

Coexistence, hysteresis, energy quantization, and competition between single- and multipulse states are common in mode-locked lasers \cite{renninger2010area,komarov2013competition,chowdhury2018multipulse,lyu2017multipulse}. Experiments and theory show that the transition from $N$ to $N+1$ pulses can pass through oscillatory and chaotic intermediate states, and that a new pulse grows from small perturbations once the saturated gain exceeds their loss \cite{bale2009transition,li2010geometrical}. A single-pulse solution may persist even where a noisy preparation usually forms several pulses. We therefore distinguish stationary structure, deterministic persistence, and stochastic accessibility. A trajectory that retains its pulse number supplies evidence for persistence; attraction additionally requires recovery after a perturbation. Endpoint fractions measure accessibility under a stated preparation and observation time.

Optical injection offers a way to probe and influence this distinction. Continuous-wave injection can create or annihilate pulses, and weak perturbations can switch between coexisting DS solutions \cite{korobko2023birth,bao2015switching}. These observations belong to a broader body of injection experiments. In passively mode-locked quantum-dot semiconductor lasers, continuous-wave (CW) injection locks one longitudinal mode, narrows the optical spectrum, and suppresses the waveform instability that appears at high pump current. Stronger injection halts mode locking in favor of injection-locked single-mode emission, with bistability and hysteresis between the two regimes \cite{rebrova2010stabilization,rebrova2011optically,habruseva2011dynamics,habruseva2012bistable,robertson2018externally}. The locking boundaries resemble those of injected single-mode semiconductor lasers \cite{wieczorek2005dynamical}, and CW injection also suppresses supermode noise in harmonically mode-locked lasers \cite{quinlan2007supermode}. In soliton fiber lasers, resonant CW injection at the milliwatt level can impose harmonic mode locking, stabilize it below a critical injected power, and add or remove single solitons \cite{niang2014harmonic,komarov2014nature,semaan2017study,ribenek2021pulse,korobko2023polarization}. The switching in Ref.~\cite{bao2015switching} was triggered by an amplitude-modulated CW that modulates the gain; such injection can also drive the internal motion of soliton molecules \cite{zou2025resonantly}. In other schemes, the injected field creates the pulse: seed pulses start Mamyshev oscillators and lower mode-locking thresholds \cite{liu2017megawatt,yang2022modelocking}, and a coherent drive sustains cavity solitons in passive and sub-threshold active resonators \cite{obrzud2017temporal,englebert2021temporal}. In our approach, the injected field is much weaker than the circulating pulse and does not initiate mode locking by itself. To our knowledge, weak injection has not yet been studied experimentally in the DSR regime. The main numerical observation is a two-stage change in the energy dependence: small injection broadens the interval dominated by single-pulse operation, whereas stronger injection produces alternating enhancement and suppression of multipulsing over an extended range. We first establish this dependence from the energy scans, then use deterministic calculations to examine whether the seed changes the soliton constituents or their formation and arrangement.

Thermodynamic language suggests useful structural comparisons, but a driven laser has no established equilibrium free energy that selects its pulse number \cite{kalashnikov2026toward}. Statistical theories of mode locking and optical wave turbulence address related but distinct questions about ordering and coherence \cite{picozzi2014wave,gordon2002phase,gat2004statistical}. The adiabatic DS spectrum supplies a core-to-cutoff scale ratio and entropy-like spectral coordinates \cite{kalashnikov2024thermodynamics,kalashnikov2025energy}. We measure these coordinates for the saved fields, separating changes of the component envelope from interference between pulses. Their predictive value for selection must be tested rather than assumed.

Nonmonotonic endpoint fractions also invite comparison with stochastic resonance (SR). Canonical SR is diagnosed by an optimum response to a time-dependent input as noise varies, with synchronization or rate matching providing additional evidence \cite{RevModPhys.70.223,anishchenko1999noise,mcdonnell2009what}. Noise-enhanced stability and inverse SR involve different observables, such as residence time or activity \cite{agudov2001noise,bacic2020inverse}. Our seed is unmodulated along the round-trip coordinate. Accordingly, we report energy-dependent formation and suppression of multipulse endpoints and separately test whether a noise optimum is resolved.

The main results begin with the unseeded and seeded energy scans, whose comparison results show the extension and subsequent nonuniformity of the single-pulse operating range. Simulations without injection or noise test whether single pulses
and prepared pulse pairs can persist at the same laser parameters.
With the seed phase held fixed, we then examine whether the pulses
formed under injection retain the shapes and spectra of unseeded
solitons. A separate scan with a fluctuating seed phase or temporal position examines
how the operating energy and additive-noise strength affect
multipulse formation. Section~\ref{sec:model} specifies the model and distinguishes the simulation protocols; Sec.~\ref{sec:results} presents the energy dependence and its supporting calculations; Sec.~\ref{sec:thermodynamic_interpretation} develops the spectral interpretation. The appendices give the analysis methods. The Supplementary Material includes the timing convention, illustrative dynamical trajectories, and potential tests of the underlying dynamics.

\section{Model and numerical experiments}
\label{sec:model}

We use the cubic--quintic complex Ginzburg--Landau equation (CGLE) in the form \cite{malomed1990kinks,cross1993pattern,kalashnikov2025energy}
\begin{align}
\frac{\partial a}{\partial z}={}&-\sigma a+(\alpha+i\beta)\frac{\partial^2a}{\partial t^2}
-i\gamma |a|^2a \nonumber\\
&+\kappa\left(1-\zeta |a|^2\right)|a|^2a+S_n(t)+\Gamma(z,t),
\label{eq:cgle}
\end{align}
where $a(z,t)$ is the slowly varying field envelope, $t$ is the local time, and $z$ counts cavity evolution. The coefficients $\sigma$, $\alpha$, and $\beta$ describe saturated net loss, spectral filtering, and group-delay dispersion, respectively. The coefficients $\gamma$, $\kappa$, and $\zeta$ describe self-phase modulation (SPM), self-amplitude modulation (SAM), and SAM saturation. We consider $\beta>0$, corresponding to NGD and the DSR regime studied in Ref.~\cite{kalashnikov2025energy}. $S_n(t)$ represents weak optical injection, and $\Gamma(z,t)$ represents the stochastic field forcing specified below.

The saturated loss depends on the total field energy $E_n$ in the simulated temporal window, including any background. Here, $n$ is the round-trip number; $z$ approximates this evolution when changes per trip are small:
\begin{equation}
\sigma_n=\vartheta\left(\frac{E_n}{E_{\rm cw}}-1\right),
\qquad
E_n=\int_{T_{\rm win}} |a_n(t)|^2dt,
\label{eq:saturation}
\end{equation}
The coefficient $\vartheta$ describes the gain-saturation stiffness. The control energy $E_{\rm cw}$ is defined by zero saturated linear net loss, $\sigma_n=0$. This is an approximate expression for the energy of a driven CW solution. Nonlinear absorption, filtering, and coherent injection also enter the field balance. We use
\begin{equation}
E'=\kappa E_{\rm cw}\sqrt{\frac{\zeta}{\beta\gamma}},
\qquad
C=\frac{\alpha\gamma}{\beta\kappa}
\label{eq:normalized}
\end{equation}
as the principal dimensionless control parameters.

The additive forcing perturbs both real and imaginary field quadratures. We implement it as complex Gaussian field kicks once per round trip \cite{206583,kalashnikov2025energy}. Conditional on the pre-kick cavity field, their grid covariance is
\begin{equation}
\begin{aligned}
\langle\delta a_{n,m}\delta a_{n,m'}^*\mid a_n\rangle
&=V_n\delta_{mm'},\\
\langle\delta a_{n,m}\delta a_{n,m'}\mid a_n\rangle&=0.
\end{aligned}
\label{eq:qnoise}
\end{equation}
Here $m,m'$ label temporal samples and $n$ labels round trips. Fresh independent Gaussian variates are used at each trip; the conditional variance depends on the current saturated net loss. In the implemented convention,
\begin{equation}
V_n=2\pi w W_q^{(\nu)}(\sigma_n),\qquad
W_q^{(\nu)}(\sigma_n)=\frac{2h\nu_0|\sigma_n|}{T_{\rm cav}}.
\label{eq:qnoise_power}
\end{equation}
The dimensionless $w$ is the scanned noise multiplier. The factor $2\pi$ follows from the source's use of $h\omega_0$, with $\omega_0=2\pi\nu_0$, rather than $\hbar\omega_0$. We retain that convention to describe the simulations actually performed. It is a phenomenological field-noise prescription, not a separately calibrated quantum-noise floor: quantum models require the gain and loss reservoirs to be specified \cite{haus2000quantum}. Appendix~\ref{app:noise} gives the increment and its normalization. The symbol $\Gamma$ in Eq.~(\ref{eq:cgle}) arises from these round-trip kicks, rather than from an independently specified continuum white-noise discretization.

\subsection{Optical injection with timing and phase fluctuations}
\label{subsec:seed_jitter_model}

The reference seed envelope is
\begin{equation}
S_0(t)=\sqrt{\frac{4A}{\pi}}\cos\!\left(\frac{\pi t}{2T_s}\right).
\label{eq:seed0}
\end{equation}
Here $A$ is the input seed-power parameter, the peak power is $4A/\pi$, and $T_s$ is the envelope time scale. Both timing and phase jitter enter the imposed field through
\begin{equation}
S_n(t)=S_0(t-\tau_n)e^{i\phi_n}.
\label{eq:seed_jitter}
\end{equation}
The delay $\tau_n$ shifts the seed envelope relative to the cavity time reference, while $\phi_n$ is its optical phase relative to the cavity field. This form permits simultaneous fluctuations in both quantities. The reported timing-jitter comparison varies $\tau_n$ with fixed phase; the phase-jitter comparison varies $\phi_n$ with fixed timing. The field ensembles, referred to below as the fixed-phase ensembles, fix both seed phase and position. These seed contributions differ from the additive complex field noise in Eq.~(\ref{eq:qnoise}), which perturbs both field quadratures.

The time-jitter in the source is defined by an independent delay on every round trip:
\begin{equation}
\begin{aligned}
\tau_n&\sim\mathcal U\!\left[-\frac{T_{\rm win}}{2q_\tau},\frac{T_{\rm win}}{2q_\tau}\right],\\
\sigma_\tau&=\frac{T_{\rm win}}{\sqrt{12}\,q_\tau}.
\end{aligned}
\label{eq:timerms}
\end{equation}
Here $q_\tau>0$ is the dimensionless ratio of the simulation-window width to the full allowed delay interval. The factor $\sqrt{12}$ converts the width of a uniform distribution to its standard deviation. Here $\sigma_\tau$ denotes the rms timing displacement of the injected seed envelope relative to the cavity reference and has units of time. It sets the spread of the independent delays, not a cumulative timing diffusion (Supplementary Sec.~\ref{supp:historical}).

For phase diffusion at zero detuning,
\begin{equation}
\phi_{n+1}=\phi_n+\delta\phi_n,
\label{eq:phasejitter}
\end{equation}
where $\delta\phi_n$ are independent zero-mean Gaussian increments. For a Lorentzian linewidth $\Delta\nu_L$ \cite{Henry1982,paschotta2004noise},
\begin{equation}
\langle\delta\phi_n^2\rangle=2\pi\Delta\nu_L T_{\rm cav}.
\label{eq:phaserms}
\end{equation}
At fixed delay, the phase factor leaves the seed intensity unchanged, $|S_n(t)|^2=|S_0(t-\tau_n)|^2$, but changes its interference with the cavity field. Timing jitter changes the envelope sampled by that field. For small delays, the forcing change is $-\tau_n S'_0(t)$ and is weak for a nearly uniform envelope. Thus, envelope timing, optical phase, and additive amplitude--phase noise represent three different perturbations of the laser dynamics.

We assume the energy--noise scan is performed by changing the cavity period $T_{\rm cav}$. The reference values are $\tau_n=0$, $T_s=T_{\rm cav}=0.8~\mu$s (for $E^{\prime}$=10), and $\Delta\nu_L=3$~MHz, giving an rms phase increment of $3.88$~rad per trip. The seed envelope is nearly uniform over $T_{\rm win}=655.36$~ps.

\subsection{Numerical experiments and state classification}
\label{subsec:numexp}

The Cr$^{2+}$:ZnS chirped-pulse oscillator parameters in Table~\ref{tab:parameters} provide the physical reference \cite{Rudenkov,kalashnikov2025energy}. At fixed average power, a change in cavity period would change the available energy.

\begin{table}[!tbp]
\caption{Parameters used for the Cr$^{2+}$:ZnS CPO simulations.}
\label{tab:parameters}
\begin{ruledtabular}
\begin{tabular}{lc}
Central wavelength $\lambda_0$ (noise-scan source) & $2.35~\mu$m \\
SPM coefficient $\gamma$ & $5.1~\mathrm{MW}^{-1}$ \\
Filter bandwidth $\Delta\lambda$ & $200$ nm \\
SAM coefficient $\kappa$ & $1~\mathrm{MW}^{-1}$ \\
SAM saturation $\zeta$ & $2~\mathrm{MW}^{-1}$ \\
Gain stiffness $\vartheta$ & $0.04$ \\
Temporal samples $N_t$ & $2^{16}$ \\
Physical time cell $\Delta t$ & $10$ fs \\
Representative seed linewidth $\Delta\nu_L$ & $3$ MHz \\
Cavity round-trip time $T_{\rm cav}$ at $E^{\prime}=$10 & $0.8~\mu$s \\
Phase rms $\sigma_\phi$ at $T_{\rm cav}=0.8~\mu$s & $3.88$ rad \\
Seed-envelope time scale $T_s$ (noise scan) & $0.8~\mu$s \\
\end{tabular}
\end{ruledtabular}
\end{table}

The simulations realize a kicked cavity map: the seed and additive noise are applied once per round trip, followed by deterministic split-step propagation with saturated loss held fixed within that trip. Each stochastic realization starts from the seed profile plus independent complex Gaussian initial noise. The multiplier $w$ scales both this initial noise and the continuing additive kicks. The resulting probabilities therefore describe formation and endpoint selection, rather than escape from a pre-existing soliton.

We denote the probability of two or more pulses by
$\mathcal{P}_2$. In
Figs.~\ref{fig:unseeded_energy_scan} and
\ref{fig:seeded_energy_scans}, each final temporal field
profile was inspected visually under the control of a final pulse profile with a constant energy level and a regular spectrum. These states were identified as containing
either one pulse or several localized pulses. These two
outcomes exhaust the realizations in those scans, so
$\mathcal{P}_2=1-\mathcal{P}_1$, where $\mathcal{P}_1$ is
the single-pulse probability.

The energy scans aim to compare unseeded and seeded dynamics. Their reference points support qualitative trends; that is, their crossovers and individual extrema are descriptive. In the stochastic calculations, endpoints are recorded after $10^4$ round trips. We performed 32 independent stochastic simulations
at each of 26 selected combinations of the control energy $E'$
and seed power $A\in\{1,10\}$~mW. The additive noise had its
reference strength ($w=1$), while the seed phase and position
were fixed. The final complex field was saved for each of the
832 simulations. We specify preparation, reference calculations, extraction, and sampling details in Appendix~\ref{app:numerics}.

\section{Results}
\label{sec:results}

\subsection{Expansion and nonuniformity of the single-pulse energy range}
\label{subsec:principal_energy_scans}

\begin{figure*}[t]
\centering
\includegraphics[width=0.5 \textwidth]{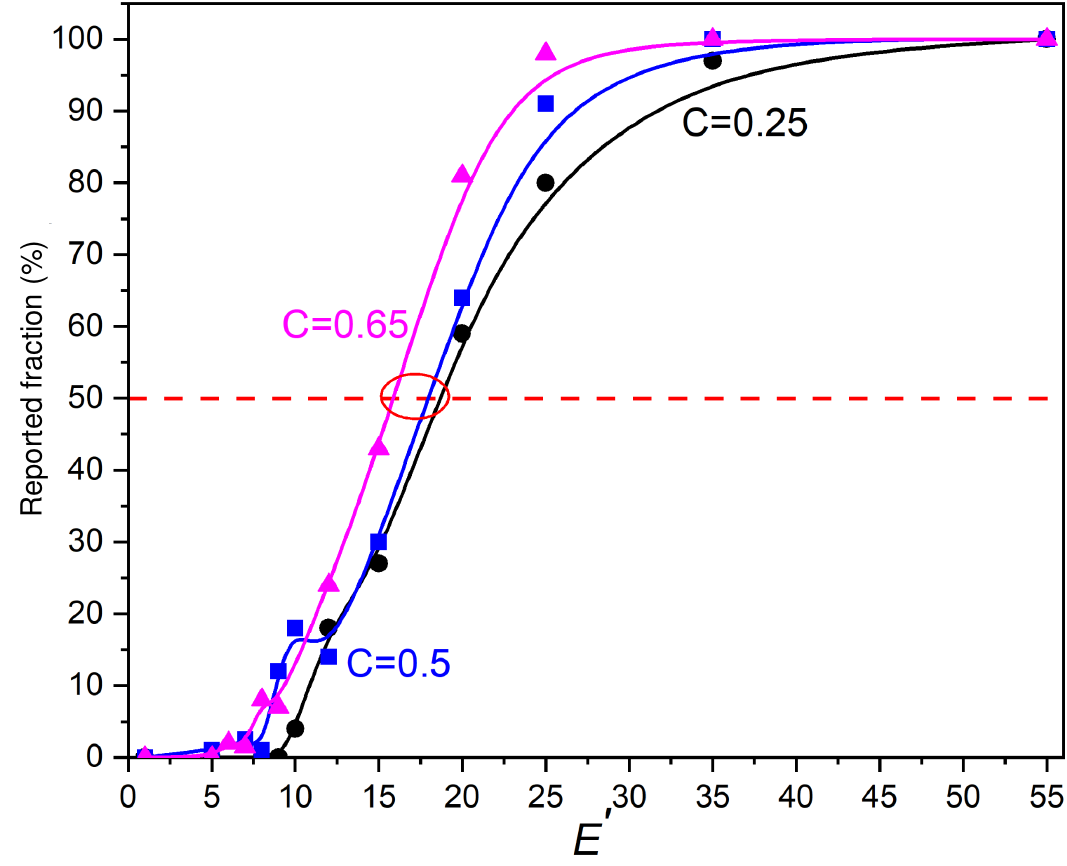}
\caption{Multipulse probability without optical
injection, determined by visual inspection of the final
temporal field profiles. Black circles, blue squares, and
magenta triangles denote $C=0.25$, $0.5$, and $0.65$,
respectively. Each realization contains either one pulse
or several localized pulses. Increasing energy favors
multipulse formation and limits predominantly single-pulse
operation. The dashed line marks
$\mathcal{P}_2=\mathcal{P}_1=1/2$ for the specified preparation
and observation time.}
\label{fig:unseeded_energy_scan}
\end{figure*}

The unseeded energy scans in Fig.~\ref{fig:unseeded_energy_scan} show the DSR limits that optical injection is intended to overcome. As the control energy increases, the fraction of multipulse outcomes rises from near zero toward unity for all three values of $C$. Predominantly single-pulse operation is therefore confined to a finite part of the scanned energy range that confines the DSR range. This statement concerns dissipative-soliton formation from an initial noise (so-called mode-locking self-start), not an upper existence boundary for the deterministic single-pulse solution.

\begin{figure*}[t]
\centering
\includegraphics[width=0.49\textwidth]{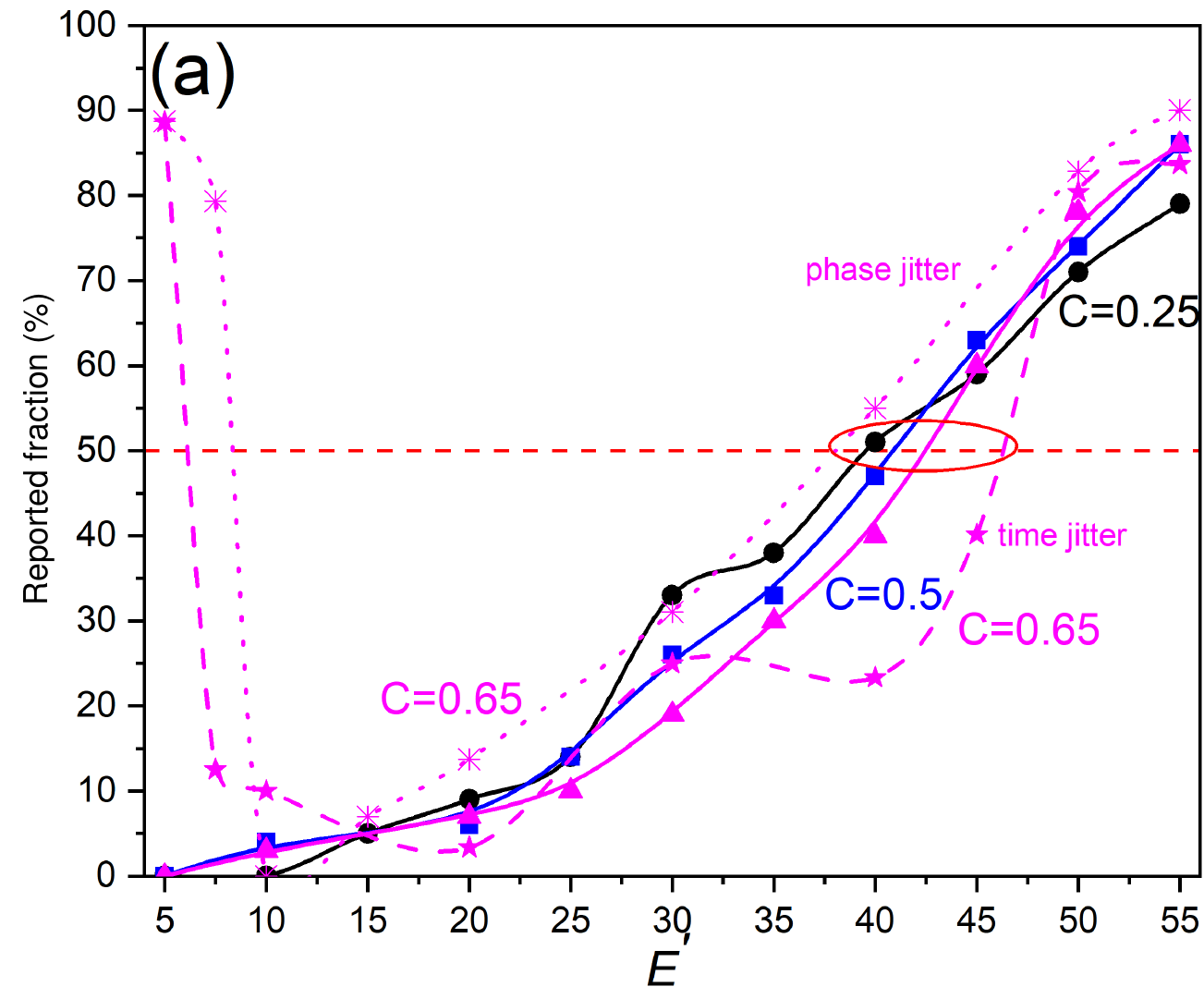}\hfill
\includegraphics[width=0.49\textwidth]{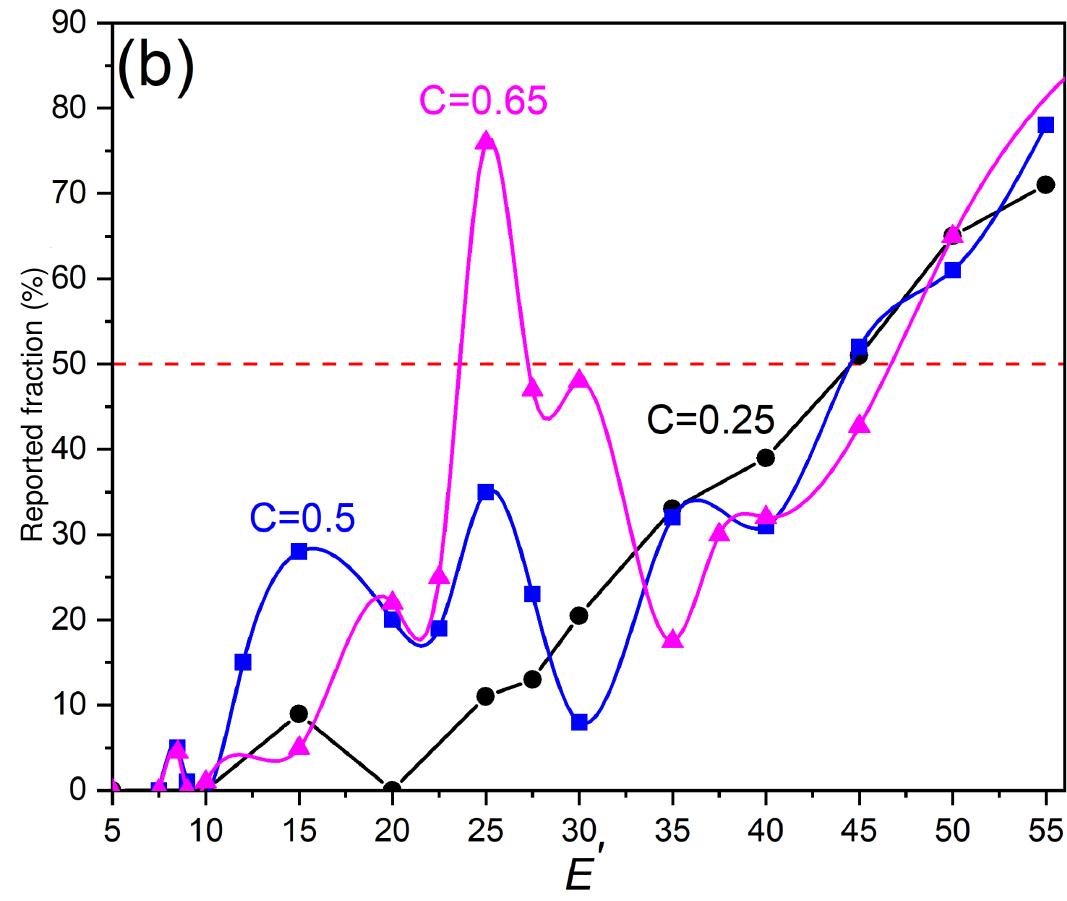}
\caption{Multipulse probability with optical injection:
(a) $A=1$~mW and (b) $A=10$~mW. All probabilities were
determined by visual inspection of the final temporal field
profiles, distinguishing one pulse from several localized
pulses. Colors and primary symbols have the meaning of
Fig.~\ref{fig:unseeded_energy_scan}. Magenta stars and
asterisks in (a) denote timing and phase jitter, respectively.
Weak injection extends predominantly single-pulse operation
toward higher energies; stronger injection introduces
alternating intervals of enhanced and suppressed multipulsing.
The dashed lines mark
$\mathcal{P}_2=\mathcal{P}_1=1/2$.
The separate ensemble with saved complex fields is presented
in Fig.~\ref{fig:seeded}.}
\label{fig:seeded_energy_scans}
\end{figure*}

Adding a small injected field substantially changes this dependence [Fig.~\ref{fig:seeded_energy_scans}(a)]. Multipulsing is suppressed relative to the unseeded curves, and the rise toward comparable single- and multipulse fractions is displaced to higher energy. Weak injection thus extends the energy interval over which the laser predominantly forms a single pulse (i.e., the DSR range).

Further increasing the seed power changes the character of this extended interval [Fig.~\ref{fig:seeded_energy_scans}(b)]. The dependence becomes nonmonotonic, particularly at $C=0.5$ and $0.65$: regions of enhanced multipulse formation are separated by regions in which it is suppressed. At $C=0.65$, the pronounced intermediate-energy maximum is followed by a return to predominantly single-pulse outcomes before multipulsing rises again at higher energy. The single-pulse operating range has become nonuniform, rather than simply shifting farther with increasing seed power.

This distinction is central to interpreting optical control. Low seed power broadens single-pulse operation, whereas stronger injection makes it more energy-selective. The following calculations examine how this selection relates to persistent single- and multipulse structures, pulse-shape changes, and noise. The statistical precision of individual maxima and
minima depends on the number of independent realizations
at each energy, as discussed in Appendix~\ref{app:wilson}.

Seed timing and phase fluctuations also modify the reported energy dependence. In Fig.~\ref{fig:seeded_energy_scans}(a), timing jitter shifts the main high-energy rise toward larger $E'$, whereas the phase-jitter curve lies above the fixed-seed curve over much of the intermediate range.

The upturn in both jittered curves at the lowest energies raises another question: has a localized pulse formed at all? In this low-energy limit, the fields also contain CW-dominated endpoints (Sec.~\ref{subsec:seeded}). We conjecture that this feature is not yet identifiable as jitter-induced multipulsing or a jitter-specific lower boundary, but may instead reflect the threshold nature of mode-locking. Experiments on injected quantum-dot lasers show a related competition: sufficiently strong CW injection halts mode locking in favor of injection-locked single-mode emission \cite{rebrova2010stabilization,habruseva2011dynamics,robertson2018externally}. A small plotted multipulse fraction alone does not establish single-pulse operation when background outcomes are possible. The extension to higher energies and the nonuniform response at stronger injection remain the main trends. The low-energy formation boundary requires a separate classification.

\subsection{Persistence of the deterministic single-pulse branch}
\label{subsec:single_persistence}

The fully deterministic scan gives a single-pulse endpoint at all 165 tested parameter points. Late-time energy fluctuations and drift are negligible over the recorded interval. This establishes that the dissipative soliton formed from the localized initial condition persists across the scan and provides a reference for the multipulse calculations.

\begin{figure*}[t]
\centering
\includegraphics[width=0.98\textwidth]{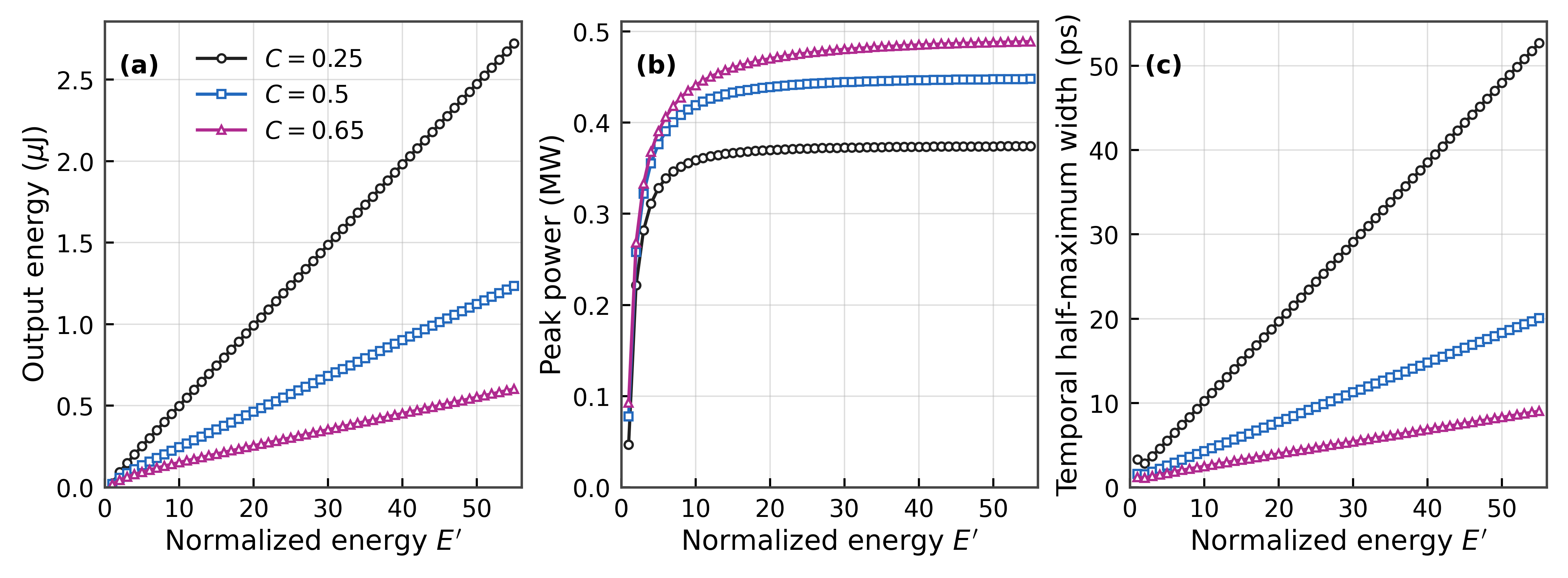}
\caption{Fully deterministic results for DS with $A=0$, $w=0$, no seed-phase or timing process, and $10^4$ round trips. Every one of the 165 parameter points in Eq.~(\ref{eq:deterministic_grid}) gives an energy-stationary single-pulse endpoint from the fixed single-pulse initial state. Shown are (a) final output energy, (b) final peak power, and (c) temporal half-maximum width. Black circles, blue squares, and magenta triangles denote $C=0.25$, $0.50$, and $0.65$, respectively. Lines connect deterministic scan points and do not represent statistical fits.}
\label{fig:deterministic_control}
\end{figure*}

Figure~\ref{fig:deterministic_control} shows the characteristic energy-scaling (DSR) mechanism: increasing the control energy broadens the pulse while its peak power changes only weakly. This provides the single-pulse reference against which multipulse configurations can be compared.

\subsection{Persistent deterministic two-pulse candidates}
\label{subsec:multipulse_discovery_results}

All 216 two-pulse initial conditions retain two localized pulses and satisfy the criteria of Appendix~\ref{subsec:multipulse_discovery_method}. Changing their initial total energy or relative phase within the tested set does not destroy the pair. Both pulses are initialized with equal energies. These simulations therefore show that an equal-energy pair can persist, but do not establish whether equal energy sharing is stable. To test this,
one must slightly increase the energy of one pulse and decrease
that of the other, keeping the total energy fixed, and observe
whether the difference decreases or grows.

\begin{figure*}[t]
\centering
\includegraphics[width=0.98\textwidth]{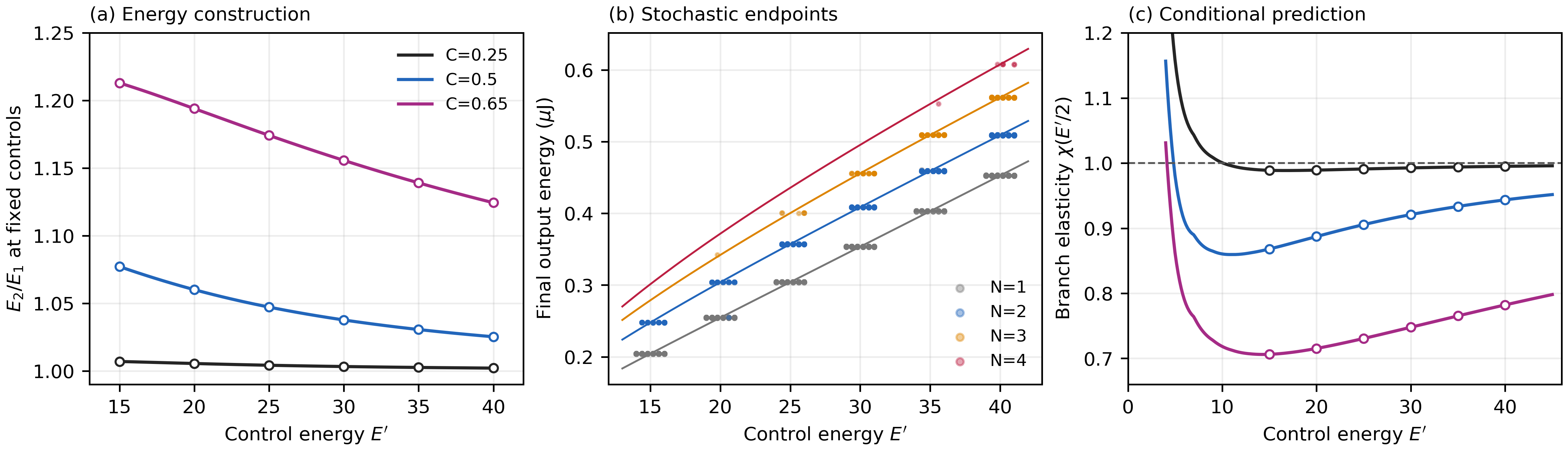}
\caption{Deterministic $N$-pulse generation. (a) Total output energy of the 216 two-pulse samples relative to the single-pulse ones at the same $(C,E')$ and the prediction $2E_1(E'/2)/E_1(E')$ of Eq.~(\ref{eq:npulse_identity}), computed from the single-pulse control alone (lines). (b) Final output energies of all 10,788 trajectories of the noise scan ($C=0.65$, $A=10$~mW) horizontally offset by noise level for visibility; lines show $NE_1(E'/N)$ for $N=1$--4 from the deterministic case. (c) Single-pulse elasticity $\chi$ of Eq.~(\ref{eq:elasticity}) evaluated at the component control point $E'/2$; symbols mark the resulting points. The line $\chi=1$ marks the threshold for conditional energy reduction, Eq.~(\ref{eq:sharing_rates}).}
\label{fig:multipulse_ladder}
\end{figure*}

Pulse pairs initialized at two different separations largely
retain their initial spacing during the simulation. This is
consistent with weak interaction over the observation time.
These results alone do not establish that either spacing is
a preferred separation of a bound pair. Such binding would
require a restoring response after a small perturbation
of the pulse separation.

At each operating point, the final energy is nearly independent of the tested separation and relative phase. A simple component construction explains this. For $S=\Gamma=0$, identical well-separated pulses share the same saturated gain. Each therefore experiences the single-pulse loss at the energy $E_{\rm cw}/N$, giving
\begin{equation}
E_N^{(0)}(E')=N\,E_1\!\left(\frac{E'}{N}\right).
\label{eq:npulse_identity}
\end{equation}

Figure~\ref{fig:multipulse_ladder}(a) compares the measured ratio $E_2/E_1$ with the prediction of Eq.~(\ref{eq:npulse_identity}) computed from the single-pulse control sample alone: the deterministic two-pulse energies fall on the predicted curves for all three values of $C$. Shared-gain pulse-number coexistence is consistent with earlier DSR studies \cite{komarov2013competition} and with the concept of DS energy quantization \cite{kalashnikov2025energy}.

At the same values of $E'$ and $C$, a pulse pair can store
more energy than a single pulse. In
Fig.~\ref{fig:multipulse_ladder}(a), the predicted total energy
of the pair exceeds the single-pulse energy by less than $1\%$
for $C=0.25$ and by up to $21\%$ for $C=0.65$.
When comparing these predictions with simulations, any
continuous background must be accounted for separately
from the energy of the localized pulses.

\begin{figure*}[t]
\centering
\includegraphics[width=0.98\textwidth]{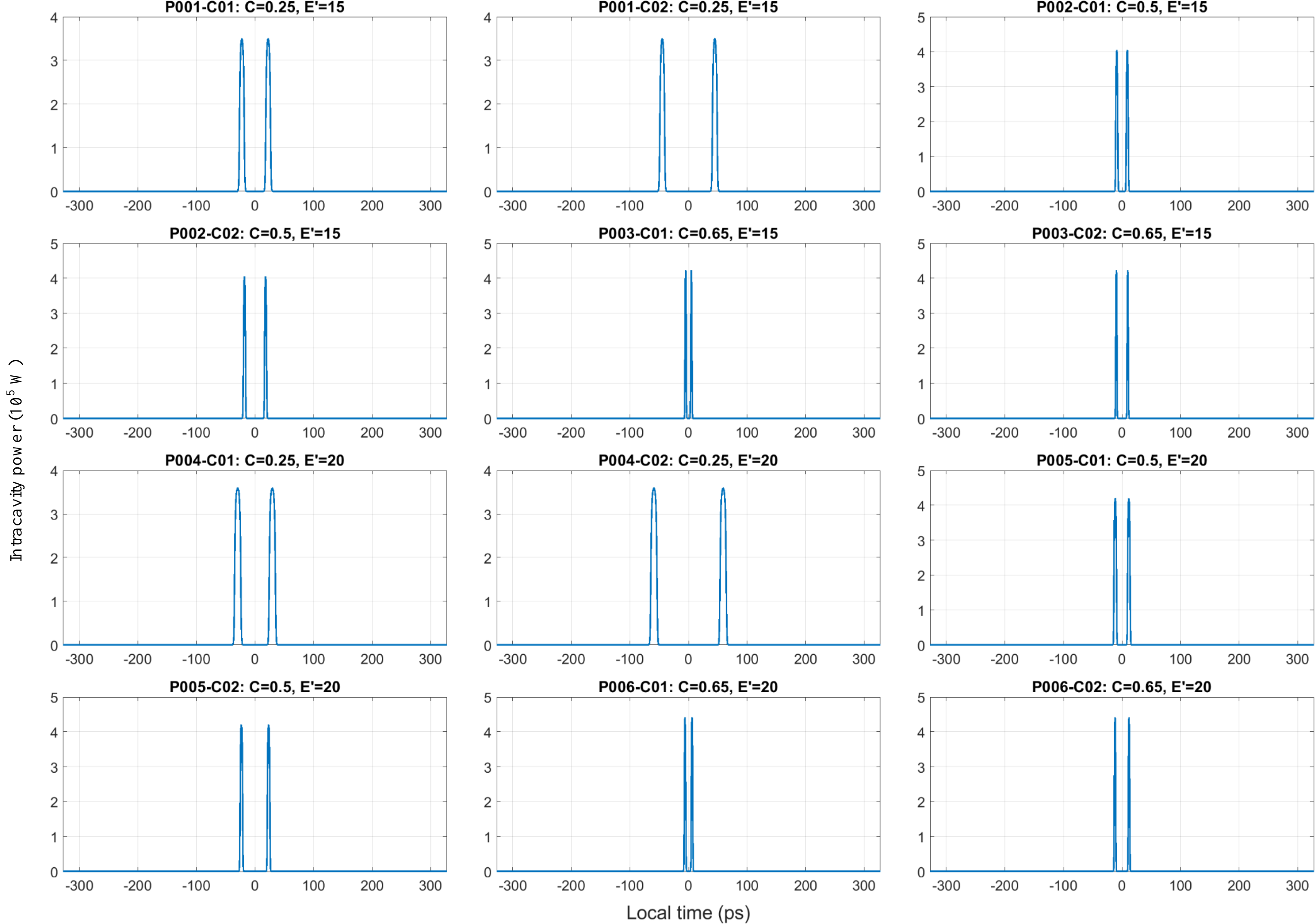}
\caption{Representative endpoint intensity profiles of the two clusters for $E'=15$ and 20 and all three values of $C$. Labels C01 and C02 denote the clusters reached from $d/T_{1/2}=3$ and 6, respectively. Two well-localized peaks support the final pulse-count classification. The retained separation classes are consistent with a separation-parameterized family, but neither the displayed intensity profiles nor the cluster labels alone establish attractive binding or complete complex-field stationarity.}
\label{fig:multipulse_profiles}
\end{figure*}

Figure~\ref{fig:multipulse_profiles} shows pulse pairs initialized
at separations $d=3T_{1/2}$ and $d=6T_{1/2}$, where $T_{1/2}$
is the intensity full width at half maximum of the reference
single pulse. The pairs survive throughout the simulations
without optical injection or noise. A separately prepared
single pulse also persists at the same laser parameters.
Thus, over the simulated time, both one-pulse and two-pulse
configurations are possible, depending on the initial field. In the ideal noninteracting limit, each pulse may be translated or phase-shifted independently. At finite separation, only global translation and phase are exact continuous symmetries of the unforced model.

\subsection{Fixed-phase endpoints and the low-energy background}
\label{subsec:seeded}
\label{subsec:windows}

The documented fixed-phase ensemble tests the seed dependence at selected energies for $C=0.65$. It starts from the broad seed envelope plus additive noise, rather than a soliton-like initial condition. At 1~mW and $E'=5$, and at 10~mW and $E'=5$, 7.5, and 8.5, the saved endpoint diagnostics indicate a dominant CW background in every realization. The spectrum is concentrated near a single line, and the field energy is well below that of the unseeded single-pulse reference. Several shallow maxima on such a background do not establish several localized solitons. These cases are excluded from the constituent comparisons and from Fig.~\ref{fig:seeded}; their diagnostic separation is discussed in Appendix~\ref{app:field_scan}.

A nearly uniform seeded start must develop temporal structure before a pulse forms. This connects the low-energy observations to the self-starting problem of mode locking \cite{haus1991selfstarting,gordon2006selfstarting}, but it does not identify the mechanism of the present driven map. At 1~mW and $E'=10$, the endpoints have single-pulse energies and spectra, while their energy traces still fail the late-time stationarity check. This is consistent with incomplete relaxation; endpoint summaries alone do not measure the formation time. Longer trajectories must distinguish slow formation from persistent driven CW operation. Energy stationarity alone cannot establish phase locking, and the lower formation boundary cannot be inferred from an unforced CW growth rate. The background fills the simulated window, so its behavior also depends on the window and gain normalization.

At the previously identified enhancement point $E'=25$, stronger injection produces a larger multipulse-label fraction. The retained counts support this pointwise difference (Appendix~\ref{app:windows}), and both counting procedures agree there. It should be noted that the automatic pulse count is not always reliable. At $E'=15$
with 10~mW injection, some fields are classified as two pulses,
although a separate analysis identifies only one localized
region with an energy close to that of a single-pulse reference.
This suggests that the number of pulses has been overestimated.
The opposite error can occur when two pulses lie close together:
the procedure may fail to distinguish them and count them as
one. Figure~\ref{fig:seeded} retains the original counts as a classifier result and flags conspicuous discrepancies. Its intervals quantify sampling uncertainty only.

The fixed-phase comparison supports enhanced multipulse formation in part of the intermediate-energy range. The finite sample size and limitations of automatic
counting concern this separate ensemble, shown in
Fig.~\ref{fig:seeded}. The probabilities in
Figs.~\ref{fig:unseeded_energy_scan} and
\ref{fig:seeded_energy_scans} were obtained from inspected soliton profiles in separate simulations. The broader physical pattern, i.e., extending single-pulse operation with weak injection, followed by a nonuniform energy response at higher seed power, appears in Figs.~\ref{fig:unseeded_energy_scan} and \ref{fig:seeded_energy_scans}. The field analysis below examines the soliton structures and arrangements associated with that response.

\begin{figure*}[t]
\centering
\includegraphics[width=0.98\textwidth]{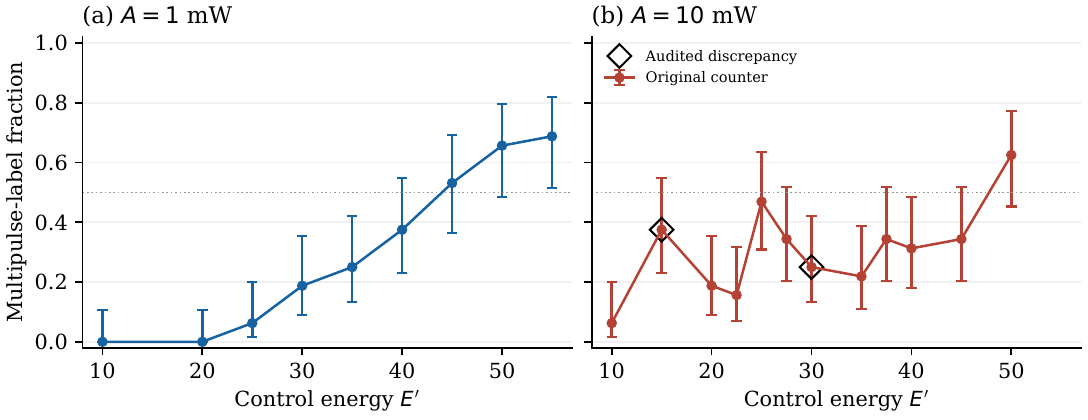}
\caption{Original multipulse fractions in the calculated fixed-phase ensemble at $C=0.65$, $w=1$, and $10^4$ round trips: (a) $A=1$~mW and (b) $A=10$~mW. Each point uses all 32 realizations and the original energy counter; bars are pointwise Wilson 95\% intervals. Diamonds flag the audited disagreements at $E'=15$ and 30 in (b), not corrected probabilities. At $E'=15$, the counter labels 12 endpoints as multipulse whereas the extraction and energy check each identify three candidates; the true count requires field inspection. Counting errors can occur in either direction. The pointwise comparison at $E'=25$ is unaffected by these identified discrepancies. The text discusses lower-energy CW-dominated cases. Lines connect samples; the dotted line marks a reported fraction of one half.}
\label{fig:seeded}
\end{figure*}

\subsection{Additive-noise scan: accessibility without a resolved optimum}
\label{subsec:noise_scan}

The energy--noise scan fixes $C=0.65$, $A=10$~mW, zero seed--cavity detuning, and a diffusive seed phase with $\Delta\nu_L=3$~MHz, and varies $E'$ from 15 to 40 and the implemented noise-power multiplier $w$ from 0 to 4 (Appendix~\ref{app:scan}).

\begin{figure*}[t]
\centering
\includegraphics[width=0.96\textwidth]{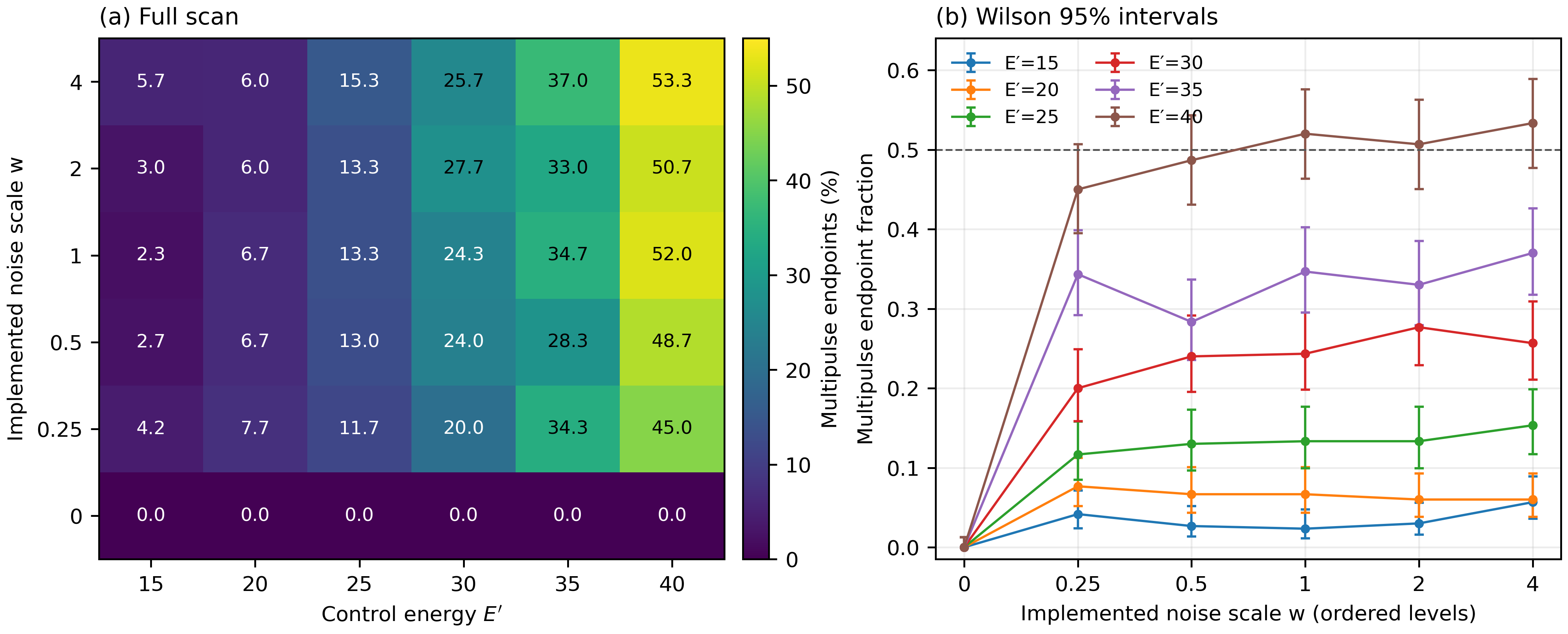}
\caption{Multipulse probability for $C=0.65$, $A=10$~mW, diffusive seed-phase jitter with $\Delta\nu_L=3$~MHz, and $10^4$ round trips. (a) Complete $(E',w)$ grid; cell values are percentages and $w$ is the implemented noise-power multiplier. (b) The same data shown against additive-noise scale, with Wilson $95\%$ confidence intervals. The horizontal dashed line marks $\mathcal{P}_2=1/2$, and connecting lines only guide the eye. Each point contains $N=300$ classified trajectories except $(E',w)=(15,0.25)$, where $N=288$. At $w=0$, the additive kicks and initial additive noise are absent, but linewidth-driven seed-phase diffusion is retained.}
\label{fig:noise_scan}
\end{figure*}

Figure~\ref{fig:noise_scan} separates the strong energy dependence from the weaker dependence on noise strength. Without additive noise, no multipulse endpoint is observed at the sampled energies, although seed-phase diffusion remains active. With additive noise, multipulse formation becomes possible throughout the grid and grows mainly with energy, approaching equal endpoint fractions near its upper end.

Grouped-binomial analysis resolves a modest common increase of multipulse probability with noise power, on top of the dominant energy trend. It finds no supported energy--noise interaction or energy-specific optimum (Appendix~\ref{app:logit}). Overlapping pointwise uncertainty intervals should therefore not be read as complete noise independence.

Multipulse formation depends much more strongly on the operating
energy than on the additive-noise strength. Over the nonzero
noise levels studied, stronger noise produces a small overall
increase in multipulsing. No clear maximum at an intermediate
noise strength is found. The sharper energy-dependent variations observed with a fixed
seed phase are not resolved in the calculations with phase
fluctuations. The noise strength is changed both in the initial field and
during its subsequent evolution. These simulations therefore
examine pulse formation, rather than the breakup of an already
prepared soliton. The final pulse counts alone do not reveal
how the additional pulses appear. The final energies also lie close to the unseeded component ladder [Fig.~\ref{fig:multipulse_ladder}(b)].

Higher pulse numbers appear at the upper energies, so $\mathcal P_2$ aggregates several endpoint classes. This phase-diffusive dataset supports the energy organization of the states, while the separate fixed-phase ensemble below tests their complex components. Any future rate description must establish whether the different multipulse arrangements can legitimately be grouped into one dynamical state.

The equal-fraction point of a descriptive fit lies near the upper energy boundary of the scan. It is an accessibility crossover for the stated preparation and observation length. Interpreting it as kinetic coexistence would additionally require stationary occupation and measured forward and reverse transitions. The corresponding rate assumptions are summarized in Supplementary Sec.~\ref{supp:rates}.

\subsection{Complex-field evidence for soliton constituents}
\label{subsec:field_identity}

The calculated field ensemble tests whether the energy ladder also describes the constituent optical fields. Each localized component is compared with a deterministic single-pulse reference at $E'/N$ and the same $C$, after background subtraction and alignment in time and global phase. The normalized complex overlap tests shape, including chirp. Energy is checked separately (Appendix~\ref{app:field_scan}).

Under the 1~mW seed, 193 extracted multipulse components reproduce the corresponding unseeded single-pulse field (squared overlap $Q>0.99$), energy, and spectral envelope. This extends the energy-sharing picture to the soliton constituents themselves. The main exception is an endpoint whose extraction may combine a close pair. These results suggest that weak injection changes the likelihood
of forming one or several pulses without substantially changing
the individual pulse shapes and spectra. The pulses within a
multipulse state remain close to unseeded solitons of comparable
energy.

At 10~mW, the single-pulse endpoints depart increasingly from the unseeded family: their median squared overlap decreases from 0.995 at $E'=10$ to 0.66 at $E'=45$. Their median overlap with references computed under the same fixed-phase drive remains at least 0.996. Stronger injection can therefore reshape a pulse as well as change its formation fraction. The comparison uses energy-stationary, extraction-eligible endpoints without close-pair flags; the anomalous driven reference at $E'=50$ is excluded.

\subsection{Component spectral scales and interference entropy}
\label{subsec:field_spectra}

A single soliton spectrum consists of a central core of width $\Xi$ and an outer edge at $\Delta$. The ratio \( r = \Delta/\Xi \) measures how narrow the core is in relation to the entire spectrum. As control energy increases along the single-pulse branch, the core narrows while the outer edge remains largely unchanged.

\begin{figure*}[t]
\centering

\includegraphics[width=0.98\textwidth]{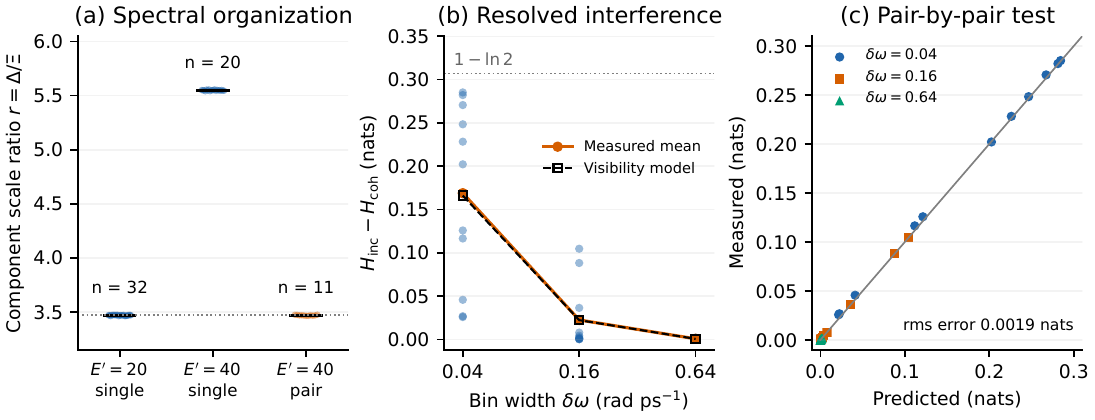}

\caption{Component structure and interference entropy at $C=0.65$, $A=1$~mW, $w=1$, with fixed seed phase and position. (a) Pulse-pair constituents recover the scale ratio of single pulses at the reduced control energy. Light markers denote trajectories and black bars their means. (b) For the same 11 pairs, the entropy difference between intensity-summed and coherent spectra decreases as the bins wash out interference. Points show individual differences, filled markers their means, and open squares the visibility-law prediction. The dotted line is the fully resolved-fringe limit. (c) Measured versus predicted differences at all three bin widths, with the equality line. Spectra use common physical bins and natural logarithms.}
\label{fig:field_spectra}
\end{figure*}

In contrast, a weakly driven pair follows a different part of the same family: each component has a spectral scale ratio comparable to that of a single soliton at a reduced energy of \( E'/2 \) [Fig.~\ref{fig:field_spectra}(a)]. Consequently, the pair comprises lower-energy solitons rather than two identical copies of the original single pulse at the initial energy \( E' \).

While agreement on the ratio \( r \) alone does not confirm that the fields are identical, the overlap comparison discussed in Section~\ref{subsec:field_identity} serves as a complementary shape test. Furthermore, Section~\ref{subsec:thermo_test} broadens the spectral comparison to the calculated ensemble.

Spectral entropy also reveals the effect of interference between
pulses. We compare the spectrum of the complete pulse pair with
the sum of the two individual pulse spectra, using the same
normalization and frequency resolution. The sum provides a
reference without interference; it does not imply that the
actual pulses have lost coherence. When interference fringes
are resolved, they concentrate spectral power into alternating
maxima and minima, lowering the entropy. Averaging over wider
frequency intervals smooths out these fringes and makes the
entropy difference smaller [Fig.~\ref{fig:field_spectra}(b)].

The pulse separation determines the fringe spacing, while the
frequency resolution determines how clearly the fringes are
seen. The expression derived in Appendix~\ref{app:fringes}
reproduces the measured entropy differences without fitting
their magnitude [Fig.~\ref{fig:field_spectra}(c)]. It assumes
that the two pulses have similar spectra and that the overall
spectral shape changes little across each fringe. Differences
between the pulse shapes and numerical sampling can affect
the agreement. The main physical result is that the entropy
difference reflects interference between the pulses; it does
not, by itself, indicate greater disorder or a loss of coherence.

\subsection{Resolving close pulses before component comparison}
\label{subsec:resolved_close_pair}

\begin{figure*}[t]
\centering
\includegraphics[width=0.98\textwidth]{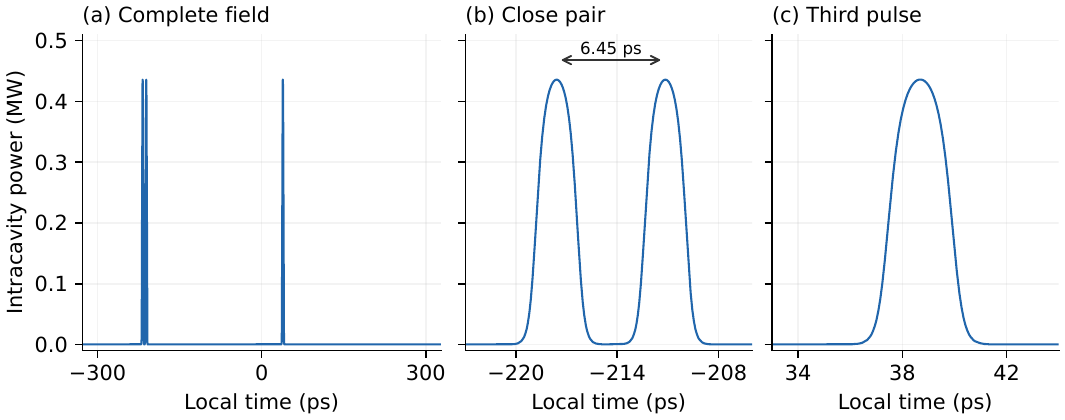}
\caption{Three intensity-resolved pulses in one fixed-phase endpoint ($C=0.65$, $E'=27.5$, $A=10$~mW, $w=1$; trajectory 610). (a) Intracavity power across the temporal window. (b) The close pair merged by automatic extraction. (c) The isolated third pulse. The pair has a deep intervening valley and a separation of 6.45~ps. Phase locking, binding, and return after perturbation require the complex fields and their evolution.}
\label{fig:resolved_three_pulses}
\end{figure*}

Figure~\ref{fig:resolved_three_pulses} shows three pulses: two close together and a third farther away. The automatic procedure merged the close pair and therefore reported only two pulses. The energy of that merged region is consistent with two soliton constituents, while the isolated region contains one. A poor match between the merged region and a single-soliton reference is therefore a counting problem, not evidence for a different soliton type. The isolated pulse matches the single-pulse reference in energy and spectral scale at $E'/3$. The merged region cannot be interpreted as one Lorentzian pulse.

The exported entropy data provide a useful screening diagnostic beyond this example. An anomalous change with bin width is consistent with interference from a close pair inside one extracted region. Such signatures are concentrated near the enhancement window in the 10~mW scan, suggesting that injection may favor particular close configurations.

\section{Spectral structure and pulse selection}
\label{sec:thermodynamic_interpretation}

The spectral measurements connect the pulse-energy sequence to the thermodynamic-like description of DSR. Their central distinction is physical: energy changes alter each soliton's spectrum, whereas arranging coherent solitons changes their interference. Both affect spectral entropy, but they describe different features of the optical field.

\subsection{Measured component branches}
\label{subsec:thermo_test}

\begin{figure*}[t]
\centering
\includegraphics[width=0.98\textwidth]{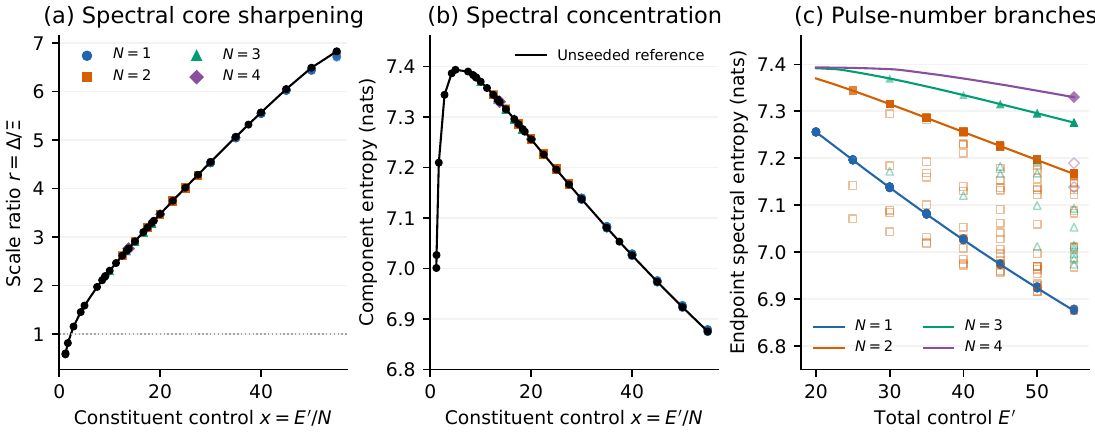}
\caption{Spectral structure at $C=0.65$. (a) Scale ratio and (b) fixed-bin entropy versus constituent control $x=E'/N$. Black dots are unseeded deterministic references; connecting lines indicate sampling. Colored symbols denote the 361 screened constituents of weak-seed endpoints, grouped by their original pulse count. The low-$x$ references are sparse; no precise thermodynamic transition is inferred from their interpolation. (c) Single-component entropy branches $H_1(E'/N)$ and the corresponding endpoint spectra: filled symbols use the component intensity sum, while open symbols show coherent multipulse spectra. The single-pulse coherent and intensity-summed values coincide. All entropies use $\delta\omega=0.04~\mathrm{rad\,ps^{-1}}$ and natural logarithms. Selection of the comparison subset is independent of agreement with the displayed entropy branches.}
\label{fig:thermo}
\end{figure*}

Figure~\ref{fig:thermo} combines deterministic single-pulse references with the weak-seed constituents. Along the reference branch, increasing control energy sharpens the spectral core relative to its outer edge, so $r=\Delta/\Xi$ increases. The fixed-bin entropy turns over in the sparsely sampled low-energy region and then decreases as the spectrum concentrates. The high-energy trend is consistent with the structural condensation picture of the adiabatic DSR theory \cite{kalashnikov2024thermodynamics,kalashnikov2025energy}. It does not imply an equilibrium temperature or a count of incoherent modes.

The multipulse constituents follow the reference branch at the reduced $x=E'/N$. Their component entropy and scale ratio closely follow those of the unseeded reference at $x$. Thus, the energy sequence also organizes the component spectra, over a subset whose selection is stated explicitly in Appendix~\ref{app:thermo_branch}.

For identical components with interference removed, the normalized spectral distribution matches that of a single constituent. Its entropy consequently satisfies the approximate relation
\begin{equation}
H_{\rm inc}^{(N)}(E')\simeq H_1(E'/N),
\label{eq:entropy_ladder}
\end{equation}
where $H_1$ is the single-pulse fixed-bin entropy. The background-subtracted, intensity-summed component spectra follow these branches [Fig.~\ref{fig:thermo}(c)]. Interference between pulses can reduce the spectral entropy
by concentrating spectral power in the fringe maxima.
The agreement with the reference calculations supports the
description of multipulse states as combinations of
single-soliton components. It does not show that the laser
selects a particular pulse configuration because of its
spectral entropy.

\subsection{What the coordinates say about selection}
\label{subsec:fragmentation}

The fixed-phase comparison provides a useful interpretive constraint. At $E'=25$, the two seed strengths give substantially different classified multipulse fractions, yet their deterministic single-pulse references differ by only about 1.3\% in $r$ and 0.004~nats in fixed-bin entropy. These scalars therefore miss part of the seed response that is visible in the full complex-field comparison. They have not established a quantitative selection law. Approximate agreement of two coordinates does not prove equality of the full competing states, nor does it rule out a sensitive or higher-dimensional dynamical description.

A spectral comparison between one pulse and $N$ components at the same control is
\begin{equation}
\Delta H_N(E')=H_1(E'/N)-H_1(E').
\label{eq:fragmentation_entropy}
\end{equation}

On the decreasing part of the single-pulse entropy branch,
sharing the energy among several pulses increases the spectral
entropy of each component. This provides a direct connection
between energy sharing and the spectral structure of multipulse
states. The resulting entropy differences offer a basis for
testing how spectral changes accompany pulse splitting and
the formation of different pulse configurations.

A complementary comparison at equal total optical energy
is obtained by choosing the component control $x$ from
$N E_1(x)=E_1(E')$. The corresponding entropy difference is
then $H_1(x)-H_1(E')$. This distinguishes the comparison at
fixed laser parameters from that at fixed total optical energy.
Additional reference spectra at low energies, together with
checks of sensitivity to the spectral cutoff, would allow
any entropy crossings to be located more reliably.

Equation~(\ref{eq:entropy_ladder}) is not an extensive entropy rule. Normalizing a sum of identical component spectral intensities recovers one component's distribution, rather than multiplying its entropy by $N$. Resolved interference introduces the additional, resolution-dependent correction measured in Sec.~\ref{subsec:field_spectra}. Averaging spectra before taking entropy also differs from averaging individual-field entropies. A thermodynamic-like comparison must therefore specify the spectral object and its resolution.

The seed changes the deterministic forcing and the formation pathway. Additive noise changes how the evolving field explores these dynamics. Describing the seed as work-like and the fluctuations as heat-like is suggestive, but it provides no first law or free energy for this cavity map. The present evidence supports dynamical selection among structurally related endpoints; a stationary occupation law would require matched initial conditions and time-resolved transitions.

\subsection{Shared gain and slow energy exchange}
\label{subsec:partition}

The common-gain reservoir links the pulses even when their fields barely overlap. Under an adiabatic approximation in which each pulse rapidly follows one shape family, the measured single-pulse energy $F(x)=E_1(x)$ defines an elasticity
\begin{equation}
\chi(x)=\frac{d\ln F(x)}{d\ln x},\qquad x=E'/N.
\label{eq:elasticity}
\end{equation}

The reduced gain-balance equations predict the ratio $1-\chi$ between the linear relaxation rate of an energy imbalance and that of a symmetric energy perturbation of the same $N$-pulse state (Appendix~\ref{app:sharing}). When $\chi$ is close to one from below, total energy can settle much faster than the energy difference between pulses. A negative imbalance relaxation rate would instead indicate growth within this approximation. Asymmetric perturbations therefore provide information that the total-energy trace alone cannot supply.

The prediction assumes that each pulse adjusts its shape much
faster than the energy imbalance between the pulses evolves,
and that their direct interaction is weak. It can be tested
by slightly increasing one pulse's energy and decreasing the
other's, then observing whether equal sharing is restored.

Small changes in pulse separation and relative phase provide
additional tests. A return to the same energy balance,
separation, and relative phase would support a stable bound
pair. If the pulses persist at different separations or phases,
their interaction may be too weak to select a unique arrangement.
Longer simulations are needed when slow evolution continues.

The slope of the single-pulse energy curve provides a prediction
for the response to an energy imbalance. Direct perturbation
tests are needed to determine where the pulse pair is stable.

\section{Discussion}
\label{sec:discussion}

Weak injection extends the energy range in which the laser
predominantly forms a single pulse. Increasing the seed power
further makes this range nonuniform: multipulsing becomes more
preferable in some energy intervals and remains suppressed in others.
The deterministic calculations show that, at the same control parameters, both a single pulse and a prepared pair can persist in the unforced model. An increased multipulse fraction in the stochastic scans therefore need not imply that the unforced single-pulse branch disappears. Whether both configurations remain attracting under a particular injected field requires a separate driven stability test. Under weak injection, the individual pulses in a multipulse state closely resemble lower-energy unseeded solitons; stronger injection can also change their shapes.

In the separate ensemble,
some low-energy realizations remain dominated by a continuous
background. Longer simulations are needed to determine
whether these fields eventually develop pulses or remain
dominated by the background under injection.
These observations concern that ensemble. In
Figs.~\ref{fig:unseeded_energy_scan} and
\ref{fig:seeded_energy_scans}, the soliton profile inspection establishes
single-pulse or multipulse formation for every realization,
including the low-energy timing- and phase-jitter results.

The closest experimental evidence comes from injected mode-locked lasers outside the DSR regime. In quantum-dot lasers, injecting a few tens of microwatts of CW into a laser emitting about 1~mW suppresses high-pump waveform instability and keeps the pulse shape stable up to about 2.4 times the threshold current, compared with 1.4 times without injection \cite{rebrova2010stabilization}. In harmonically mode-locked fiber lasers, CW injection widens the stable operating pump range \cite{korobko2023polarization}, whereas stabilization holds only below a critical injected power and at discrete wavelengths \cite{semaan2017study}. In simulations of the soliton fiber laser of Ref.~\cite{korobko2023birth}, a CW in phase with the cavity mode stabilizes the single-soliton state, whereas a phase-shifted CW can trigger the birth of a second soliton. These results parallel three features of our calculations: a weak coherent seed extends the energy range of a pulse configuration, a stronger or fluctuating seed can promote additional pulses, and a CW seed competes with mode locking. That a seed far weaker than the pulse can bias formation from noise is also familiar from injection-seeded Q-switched lasers, in which a seed many orders of magnitude weaker than the output pulse selects the axial mode from which the pulse builds up \cite{park1984single}.

The fiber experiments concern anomalous-dispersion solitons, which shed dispersive waves (Kelly sidebands) that can interfere with the injected CW \cite{komarov2014nature}. Strongly chirped DSR pulses in normal dispersion do not form Kelly sidebands, so in the present model the seed acts through the background, the saturated gain, and direct interference with the pulse. The only injection experiment on a DSR laser that we found synchronizes its repetition rate through cross-phase modulation and reports pulse splitting at higher pump without studying it \cite{pokryszka2024stabilized}. A direct test would inject a weak, narrow-linewidth CW seed into a DSR fiber laser or a Cr$^{2+}$:ZnS chirped-pulse oscillator and measure the single-pulse fraction over repeated starts as the pump is scanned, with and without imposed phase noise on the seed. In the present model, the seed carries about 0.03--1.5\% of the intracavity average power at $E'=5--45$; the quantum-dot experiments injected up to a few percent of the emitted power \cite{rebrova2010stabilization}.

The thermodynamic context is formulated in spectral terms. Sharing the energy among several pulses changes the spectrum
of each individual pulse, particularly the width of its central
peak. The separation and relative phase between the pulses
also affect the interference fringes in the combined spectrum.
Spectral entropy reflects both contributions. The measured entropy branches and fringe-visibility law explain these two contributions without favoring one pulse number as the equilibrium state. Similar values of $r$ and entropy can accompany a substantial change in endpoint fractions; that observation motivates a dynamical description but does not exclude other structural predictors or a future nonequilibrium potential.

The noise scan addresses a separate question. It shows a modest, common increase in multipulse fraction across the sampled nonzero noise levels, with no preferred finite noise level. This supports noise-assisted formation under the implemented initial conditions, not canonical SR. A direct SR experiment would add a weak periodic modulation along the round-trip coordinate and measure its coherent response as noise varies, together with transition synchronization or rate matching \cite{RevModPhys.70.223}. Amplitude-modulated CW injection already locks the switching between coexisting DS states \cite{bao2015switching} and drives soliton-molecule dynamics \cite{zou2025resonantly}, so such a test is experimentally accessible.

\section{Conclusion}
\label{sec:conclusion}

The energy scans indicate that weak injection extends single-pulse operation toward higher energies. Further increasing the seed power makes that extended range nonuniform, enhancing multipulsing in some intervals and suppressing it in others. At low energies, a continuous background can still dominate the field. Experiments on injected semiconductor and soliton fiber lasers show related effects: weak injection extends stable operation, stronger injection quenches mode locking, and injection can add or remove pulses. Testing a weak seed in a laser operating in the DSR regime remains an open experimental question.

In the unforced model, prepared single pulses and pulse pairs can persist under the same settings. Under weak injection, the observed multiple pulses largely retain the structure of lower-energy single solitons; stronger injection can reshape them. Their spectra separate changes within each pulse from interference between pulses, giving spectral entropy a clear structural meaning without making it a universal selection rule. A claim of stochastic resonance would also require measuring the response under weak periodic modulation.

\begin{acknowledgments}
This work was supported by Norges Forskningsr\aa d (Projects No.~303347, UNLOCK, and No.~326503, MIR) and ATLA Lasers AS.
OpenAI ChatGPT/Codex tools helped prepare simulation and post-processing code, organize the analysis, and edit the manuscript with Anthropic Claude assistance. Reported numerical results were derived from the MATLAB simulations using the IDUN NTNU scientific computing infrastructure \cite{sjalander+:2019epic}. The authors retain responsibility for the scientific content and its verification.
\end{acknowledgments}

\section*{Author Declarations}
\subsection*{Conflict of Interest}
The authors have no conflicts to disclose.

\subsection*{Author Contributions}
\textbf{Vladimir L. Kalashnikov:} Conceptualization, Methodology, Formal analysis, Investigation, Writing---original draft. \textbf{Alexander Rudenkov:} Conceptualization, Methodology, Writing---review \& editing. \textbf{Evgeni Sorokin:} Writing---review \& editing. \textbf{Irina T. Sorokina:} Funding acquisition, Supervision, Writing---review \& editing.

\section*{Data Availability}
The exported endpoint, component, reference-comparison, and spectral-entropy tables, along with scripts that reproduce the saved-field analyses, are available from the corresponding author upon reasonable request.

\section*{Supplementary Material}
The Supplementary Material contains illustrative dynamical traces, the timing-jitter convention, validity conditions for a rate description, the proposed ensemble-coherence measurement, and targeted tests of multipulse attraction. The separate data archive contains the exported tables and reproducible postprocessing scripts.

\appendix

\section{Numerical implementation and simulation protocols}
\label{app:numerics}
This appendix collects the numerical implementation of the model in Sec.~\ref{sec:model} and the simulation protocols summarized in Sec.~\ref{subsec:numexp}.

\subsection{Additive noise}
\label{app:noise}

The adopted source adds the following increment at each temporal sample:
\begin{equation}
\delta a_{n,m}=\sqrt{\frac{h\omega_0|\sigma_n|}{T_{\rm cav}}}
\,(\xi_{n,m}+i\eta_{n,m}),\quad \omega_0=2\pi\nu_0,
\label{eq:implemented_noise}
\end{equation}
Here $\xi$ and $\eta$ are independent unit-variance real Gaussians, giving $\langle|\delta a|^2\rangle=2h\omega_0|\sigma|/T_{\rm cav}=2\pi W_q^{(\nu)}$ at $w=1$ \cite{haus2000quantum}; a multiplier $w$ scales this variance. Since $\sigma$ depends on the field energy, the kicks are round-trip dependent.

The implementation involves adding independent complex samples to the temporal grid once per round trip. These samples are not normalized Fourier-mode amplitudes. A continuum white-noise interpretation and resolution tests must maintain a specified covariance and bandwidth despite changes in \(\Delta t\) and the FFT convention. For reproducibility, we specify the variance of the random field added at each round trip and its correlations between
temporal samples. Any dependence on the numerical grid must
be derived from this noise prescription. Scaling factors
used in other noise models require a separate normalization
check before they can be applied here.

\subsection{Kicked cavity map and initial state}
\label{app:map}

Our implementation applies the seed and additive noise once per round trip. It uses 200 symmetric split-step Fourier slices for the subsequent deterministic evolution, holding the saturated loss fixed within that trip. Thus, the reported simulations realize a kicked cavity map motivated by Eq.~(\ref{eq:cgle}).
Each shot in the energy--noise and field ensembles starts from the zero-phase seed profile plus independent complex Gaussian initial noise. In normalized units,
\begin{equation}
y_m(0)=r_0(t_m)+\sqrt{\frac{w h\omega_0\gamma\,\mathrm{OC}}{T_{\rm cav}}}
\,(\xi_m+i\eta_m),
\label{eq:initial_noise_scan}
\end{equation}
where $r_0=\sqrt{(4/\pi)\gamma A}\cos[\pi t/(2T_s)]$ and $\mathrm{OC}=0.14$ is the oscillator output-coupling factor. At $w=0$, this initial state is just the seed profile. Increasing $w$ changes both the initial random field and the ongoing additive forcing. The measured probabilities are consequently formation/endpoint-selection statistics.

\subsection{Noise-strength scan method}
\label{app:scan}

The energy--noise scan fixes $C$, $A$, zero seed--cavity detuning, and a diffusive seed phase corresponding to $\Delta\nu_L=3$~MHz. With $T_{\rm cav}=0.8~\mu$s, Eq.~(\ref{eq:phaserms}) gives an rms phase increment of $3.88$~rad per round trip, so the coherent phase is strongly decorrelated while the temporal seed envelope remains fixed. We evaluate the Cartesian grid
\begin{equation}
\begin{aligned}
E'&\in\{15,20,25,30,35,40\},\\
w=\texttt{noiseScale}&\in\{0,0.25,0.5,1,2,4\}.
\end{aligned}
\label{eq:noise_scan_grid}
\end{equation}
Each trajectory corresponds to $10^4$ round trips on $N_t=2^{16}$ temporal samples using 200 split-step slices per round trip. The classifier and all its thresholds are held fixed across the grid. The target is 300 independently seeded trajectories per point.

\subsection{Fully deterministic unseeded control}
\label{app:deterministic}

To distinguish persistence of an already formed pulse from stochastic endpoint selection, we additionally set $A=0$ and $\Gamma=0$.
We scan
\begin{equation}
C\in\{0.25,0.50,0.65\},\qquad E'=1,2,\ldots,55.
\label{eq:deterministic_grid}
\end{equation}

The deterministic scan covers 165 combinations of $C$ and $E'$.
Every simulation starts from the same initial field
$a(0,t)=\sqrt{P_0}\,\operatorname{sech}(t/T_0)$ with $P_0=1$~W and $T_0=5$~ps without injection and noise. The results therefore describe how this initial field evolves across the chosen parameter range. Because no random variable is present, one trajectory per point is sufficient to define the endpoint of this specified initial state, but it cannot define a basin volume or an endpoint probability. ``Stationary'' below denotes satisfaction of the late-time numerical classifier.

\subsection{Deterministic two-pulse discovery scan}
\label{subsec:multipulse_discovery_method}

The multipulse search uses the converged deterministic single-pulse field $a_1(t)$ at each of the 18 matched points
\begin{equation}
C\in\{0.25,0.50,0.65\},\qquad
E'\in\{15,20,25,30,35,40\}.
\label{eq:multipulse_discovery_grid}
\end{equation}
A controlled two-pulse initial state is constructed as
\begin{equation}
\begin{aligned}
a_2^{(0)}(t)={}&\mathcal{N}_q\left[
\sqrt{\eta}\,a_1\!\left(t-\frac{d}{2}\right)\right.\\
&\left.+e^{i\phi}\sqrt{1-\eta}\,
a_1\!\left(t+\frac{d}{2}\right)\right],
\end{aligned}
\label{eq:two_pulse_initial}
\end{equation}
Here $d$ is the separation, $T_{1/2}$ is the reference pulse's intensity FWHM, $\phi$ is the relative phase, $\eta$ is the energy partition, and $\mathcal{N}_q$ fixes the total initial energy to $q$ times the reference single-pulse energy. The discovery grid uses $d/T_{1/2}\in\{3,6\}$, $\phi\in\{0,\pi\}$, $\eta=1/2$, and $q\in\{0.8,1,1.2\}$, giving 12 initial states per point and 216 deterministic trajectories. The simulations use $A=w=0$, no phase or timing jitter, $N_t=2^{16}$ samples, 200 split-step slices per round trip, and $2\times10^4$ round trips.

We classify the last half of each diagnostic propagation trajectory. The state is classified as multipulse if at least two
pulses are resolved in $95\%$ or more of the recorded samples
in this interval, and the relative drift in total energy
does not exceed $10\%$. We further identify approximately stationary candidates. For these states, the standard deviations of
total energy and pulse separation must be no larger than
$3\%$ and $5\%$ of their respective mean values. The standard
deviation of the pulse-energy ratio must not exceed $0.05$.
The relative phase must also remain concentrated around a
preferred value, with
$|\langle e^{i\Delta\phi}\rangle|\geq0.90$, where the average
is taken over the same interval. These criteria provide an initial selection of states for further study.
The reported success fraction is the number of accepted runs
divided by the 12 prescribed initial states. It describes the
outcomes for this particular set of starting fields. It does
not give the probability of forming the state under random
noise or measure the full set of initial fields that evolve
toward it.

Symmetry, rather than the system dynamics, determines two of these diagnostics. When \(\eta = 1/2\), the components behave as identical copies of \(a_1\). As long as they do not interact, they experience the same saturated loss and rotate their phases at the same rate. This ensures equal energy sharing and, by design, maintains a constant relative phase. Therefore, under these conditions, the component-energy ratio and the degree of phase locking cannot differentiate between binding and non-interaction. Only initial states with \(\eta \neq 1/2\), or perturbations in separation and relative phase, can effectively test these scenarios.

\subsection{Saved-field ensemble and component comparisons}
\label{app:field_scan}

We retain the complex fields from simulations at $C=0.65$
with seed powers $A=1$ and $10$~mW. Each energy point has
32 realizations with additive noise at $w=1$, fixed seed
phase and position, and no timing or phase jitter. The
comparison uses the fields after $10^4$ cavity round trips.

Individual pulses are isolated after subtracting the estimated
background. Profiles with a dominant CW background
or unresolved pulses require direct inspection before interpreting
their components. For $N$ pulses sharing the energy equally,
the reference is an unseeded single soliton at control energy
$E'/N$ and the same $C$.

The normalized complex-field overlap measures similarity
\[
Q=\max_{\tau}
\frac{\left|\int a_j(t)a_{\rm ref}^{*}(t-\tau)\,dt\right|^2}
{\int |a_j(t)|^2\,dt\,\int |a_{\rm ref}(t)|^2\,dt}.
\]
Here $\tau$ aligns the pulses in time. A value near unity
indicates agreement in shape and phase profile, apart from
a constant phase difference. No frequency or chirp adjustment
is made. Pulse energies are compared separately because
normalization removes their overall amplitude difference.

The tabulated field energies are the intracavity energies.
Multiplication by the output-coupling fraction
$\mathrm{OC}=0.14$ gives the output energies used in
Figs.~\ref{fig:deterministic_control} and
\ref{fig:multipulse_ladder}.

\section{Statistical analysis of endpoint probabilities}
\label{app:statistics}
This appendix specifies the Monte Carlo estimators, confidence intervals, and regression models used in Sec.~\ref{sec:results}.

\subsection{Ensemble size, confidence intervals, and rare-event sampling}
\label{app:wilson}

The probabilities in Figs.~\ref{fig:unseeded_energy_scan}
and \ref{fig:seeded_energy_scans} were obtained by inspection of the final temporal field profiles. Each
realization was identified as containing either one pulse
or several pulses. Thus, $\mathcal{P}_2$ denotes the probability
of two or more pulses, and
$\mathcal{P}_2=1-\mathcal{P}_1$, where $\mathcal{P}_1$ is the
single-pulse probability. These results do not depend on
an automatic pulse-counting procedure.

For $k$ multipulse outcomes among $N$ independent realizations,
the estimated probability is
\begin{equation}
\widehat{\mathcal{P}}_2=\frac{k}{N}.
\label{eq:mc_estimator}
\end{equation}
This is the probability of observing a multipulse state
at the end of the specified simulation time.

Where the individual outcome counts are available, sampling
uncertainty is quantified using the Wilson confidence
interval~\cite{wilson1927probable}. For confidence level
$1-\alpha$,
\begin{equation}
\begin{aligned}
\mathcal{P}_{2,\pm}
&=\frac{1}{1+z^2/N}
\left[
\widehat{\mathcal{P}}_2+\frac{z^2}{2N}
\right.\\[-1mm]
&\qquad\left.
\pm z\sqrt{
\frac{\widehat{\mathcal{P}}_2
(1-\widehat{\mathcal{P}}_2)}{N}
+\frac{z^2}{4N^2}}
\right],
\end{aligned}
\label{eq:mc_ci}
\end{equation}
where $z$ is the standard-normal quantile
$z_{1-\alpha/2}$. The intervals shown in
Fig.~\ref{fig:noise_scan} use $z=1.96$, corresponding
to $95\%$ confidence.

A finite ensemble may miss rare multipulse outcomes.
Consequently, observing no such outcomes does not establish
that their probability is zero; the confidence interval
retains a positive upper bound. Away from probabilities
zero and one, sampling uncertainty decreases approximately
as $N^{-1/2}$~\cite{brown2001interval}.

For the separate energy--noise scan, the common dependence
on energy and noise strength is examined using a binomial
regression over the 30 conditions with $w>0$:
\begin{equation}
\operatorname{logit}\mathcal{P}_2
=b_0+b_EE'+b_W\log_2w,
\label{eq:binomial_model}
\end{equation}
where $\operatorname{logit}p=\ln[p/(1-p)]$.
The following subsection reports this analysis and
comparisons allowing separate effects at each noise level.

\subsection{Separating the effects of energy and noise
on multipulse formation}
\label{app:logit}

For each value of the energy parameter $E'$, we compare
simulations with different nonzero strengths of additive
noise. Each calculation is repeated many times, and we
count how often several pulses are present at the end.
This allows us to distinguish the influence of energy
from that of noise.

Energy produces the larger change: multipulsing becomes
much more frequent as $E'$ increases. Increasing the noise
has a smaller effect. When the results at all tested energies
are considered together, they show a modest increase in
multipulse formation with noise strength.

The results do not reveal a clear maximum at an intermediate
noise strength. They therefore show that noise helps
multipulse formation, without identifying a preferred
noise level. To understand how the additional pulses
appear, their development must be followed during the
simulation, beyond inspecting the final field alone.

\subsection{Fixed-phase comparison}
\label{app:windows}

The simulations show that increasing the
seed power makes multipulse formation more likely. This result supports the main physical observation:
weak injection extends the range of predominantly
single-pulse operation, but stronger injection can favor
several pulses within parts of that range. Increasing
the seed power therefore does not improve single-pulse
operation uniformly; its effect depends on the operating
energy.

\section{Thermodynamic coordinates of the stochastic problem}
\label{sec:thermodynamic_coordinates}

The thermodynamic approach describes how a driven nonlinear
system organizes the supplied energy into a coherent
dissipative structure. A soliton persists through continuous
energy exchange with its environment, while dispersion,
nonlinearity, gain, and loss determine the distribution
of that energy within the field.

The phase-space formulation developed in
Ref.~\cite{kalashnikov2019phasespace} provides a connection
between this dynamics and statistical mechanics. It describes
the field jointly in time and frequency, including the
correlations associated with chirp and coherence. In a
suitable kinetic approximation, the spectral components
can be described as interacting quasiparticles.
This description accommodates coherent solitons as well
as fluctuating fields.

Here we examine this organization through the spectral
scales, spectral entropy, and energy sharing between pulses.
The measured formation probabilities complement these
quantities by showing how frequently the different pulse
configurations are reached.

\subsection{Spectral scales and coherent condensation}

For a strongly chirped soliton in the adiabatic
NGD approximation, the normalized spectrum has the
truncated Lorentzian form\cite{kalashnikov2024thermodynamics,kalashnikov2025energy}
\begin{equation}
p(\omega)=
\frac{\Xi}{2\arctan(\Delta/\Xi)}
\frac{\Theta(\Delta^2-\omega^2)}{\omega^2+\Xi^2},
\qquad \int p(\omega)d\omega=1,
\label{eq:spectrum_shape}
\end{equation}

Here $\Delta$ is the spectral cutoff, $\Xi$ is the
width of the central spectral core, and $\Theta$ is the
Heaviside function.

This spectrum has the Rayleigh--Jeans form discussed
in Ref.~\cite{kalashnikov2019phasespace}. In a representation
where the quadratic mode energy is written as $\omega^2$,
the denominator can be expressed as
$\omega^2-\mu_{\rm sp}$, with
$\mu_{\rm sp}=-\Xi^2$. Thus, $\Xi$ supplies the spectral
analog of a negative chemical potential. As $\Xi$
decreases, spectral weight becomes increasingly concentrated
near the carrier frequency. This gives the condensation
picture a direct meaning in terms of the measured spectrum.

The ratio

\begin{equation}
r=\frac{\Delta}{\Xi}
\label{eq:scale_ratio}
\end{equation}

\noindent measures the separation between the broad spectral
edges (cutoff frequency) and the narrow central core. Their inverse widths
define two characteristic time scales. Near DSR, the
central core narrows while the outer cutoff remains finite,
so these scales become increasingly separated.
The field can remain coherent throughout this process.

The role of $r$ is therefore to characterize the
spectral organization associated with energy scaling.
The ensemble-coherence measurement described in
Supplementary Sec.~\ref{supp:coherence} provides a complementary test of
how the complex pulse profile varies between realizations.

\subsection{Spectral entropy and energy redistribution}

Spectral entropy quantifies how the optical energy
is distributed among frequency intervals. Using the same
physical bin width $\delta\omega$ for all compared fields,
we define

\begin{equation}
H_{\delta\omega}=-\sum_i P_i\ln P_i,
\qquad
P_i=\int_{\text{bin }i}p(\omega)d\omega.
\label{eq:fixed_bin_entropy}
\end{equation}

Here $P_i$ is the fraction of the spectral energy
contained in the $i$th interval. A spectrum concentrated
in shorter intervals has lower entropy than a spectrum
whose energy is distributed more broadly. This quantity
characterizes the frequency distribution of the energy,
including for a fully coherent field.

Along the high-energy single-pulse branch, the
decrease of $H_{\delta\omega}$ records the increasing
concentration of spectral energy near the carrier frequency.
Together with the growth of $r$, it provides a measurable
signature of the coherent-condensation process.
Rescaling spectra to the same width allows us to compare their
shapes independently of their physical widths. Here we keep the
frequency intervals fixed, so the entropy also records whether
the optical energy becomes concentrated in a narrower frequency
range.

When the energy is shared between several pulses,
each constituent can recover the broader central spectrum
of a lower-energy soliton. Comparing the corresponding
entropies therefore connects multipulsing to a redistribution
of energy among coherent structures. The agreement between
the measured component spectra and the single-pulse
reference branches gives quantitative support to this
thermodynamic interpretation.

A further thermodynamic description can examine
energy--entropy relations along specified families of
states. The present results
establish the spectral coordinates and component relations
on which such a description can be developed.

\subsection{Spectral fringes of coherent pulse pairs}
\label{app:fringes}

A coherent pulse pair illustrates how spectral
entropy also records the arrangement of the components.
Adding their separate spectral intensities gives the
distribution associated with their individual shapes.
Adding their complex fields additionally produces
interference fringes. Under the assumptions specified
below, resolved fringes concentrate spectral weight and
reduce the entropy relative to the sum without interference.
The measured entropy difference therefore has a physical
interpretation: it quantifies how coherent interference
changes the distribution of energy across the spectrum.

For two identical coherent components with separation $d$ and phase $\phi$, the normalized spectral density is
\begin{equation}
\begin{aligned}
p_2(\omega)&=\frac{p_c(\omega)[1+\cos(\omega d+\phi)]}{Z},\\
Z&=\int p_c(\omega)[1+\cos(\omega d+\phi)]\,d\omega.
\end{aligned}
\label{eq:fringe_density}
\end{equation}
When overlap is negligible, $Z\simeq1$. If many fringes lie under a slowly varying envelope and the bins resolve them, averaging the rapidly oscillating factor gives the asymptotic prediction
\begin{equation}
H^{(2)}_{\delta\omega}(E')\simeq
H^{(1)}_{\delta\omega}(E'/2)-(1-\ln2).
\label{eq:fringe_entropy}
\end{equation}

The constant is the fringe-phase average $\langle(1+\cos u)\ln(1+\cos u)\rangle_u=1-\ln2$. Finite bins reduce the visibility. For an envelope that varies slowly across a bin, integration of its interference factor gives
\[
V=\left|\operatorname{sinc}\left(\frac{\delta\omega d}{2}\right)\right|,
\qquad \operatorname{sinc}x=\frac{\sin x}{x}.
\]
When many fringe phases are sampled under the envelope, the corresponding entropy difference is
\begin{equation}
\begin{aligned}
H_{\rm inc}-H_{\rm coh}&\simeq\mathcal F(V),\\
\mathcal F(V)&=\left\langle(1+V\cos u)\ln(1+V\cos u)\right\rangle_u\\
&=1-\sqrt{1-V^2}+\ln\frac{1+\sqrt{1-V^2}}{2}.
\end{aligned}
\label{eq:fringe_visibility}
\end{equation}
Thus, fully resolved fringes approach the deficit $1-\ln2$, whereas washed-out fringes give $\mathcal F(V)\simeq V^2/4$. Figure~\ref{fig:field_spectra} tests this finite-resolution prediction using measured separations. The formula assumes identical coherent components, negligible temporal overlap, a slowly varying spectral envelope, and adequately sampled fringes. Unequal energies or shapes further reduce visibility; finite support, bin origin, and envelope--fringe correlations require corrections. For larger pulse complexes, the interference pattern also depends on their arrangement and phases, so the two-pulse formula is not a universal pulse-number entropy law. Averaging spectral intensities before evaluating entropy removes interference differently from averaging entropies of individual coherent spectra.

\subsection{Reconstruction of component branches}
\label{app:thermo_branch}

Figure~\ref{fig:thermo} compares the pulse energy, spectral
entropy, and ratio $r=\Delta/\Xi$ of single solitons calculated
without injection or noise. Here $x$ denotes the normalized
control energy. We include only localized pulses whose energy
has stopped changing appreciably and whose spectra are well
described by the fitting model.

These previously calculated states show how the single-soliton
spectrum changes with increasing control energy. At low energies,
however, the available points are widely spaced. The connecting
lines therefore indicate the general trend; a finer scan is
needed to locate the entropy maximum or the point where $r=1$.
Establishing whether either feature accompanies pulse breakup
also requires following the pulse evolution.

The component subset uses 1~mW endpoints with $E'\geq10$, extraction eligibility, accepted spectral fits, no flagged close region in that trajectory, and an energy-stationary endpoint label. It contains 361 components from 255 endpoints: 168 single-pulse, 70 two-pulse, 15 three-pulse, and 2 four-pulse cases. Their original pulse counts are retained. Relative differences in $r$ average $-0.19\%$ with a standard deviation of $0.22\%$; component entropy differences average $0.0006$~nats with a standard deviation of $0.0012$~nats. These are descriptive component spreads, not confidence intervals for independent samples. Mean intensity-sum residuals from $H_1(E'/N)$ are below $10^{-3}$~nats in each class. No selection on either residual is imposed.

Entropy branches in panel (c) use shape-preserving interpolation of the unseeded references, restricted to their sampled domain. Reference entropy is available at $0.04~\mathrm{rad\,ps^{-1}}$; component spectra are also exported at 0.16 and $0.64~\mathrm{rad\,ps^{-1}}$. Fixed binning permits paired comparisons but does not guarantee bin-width invariance or cancellation of all discretization effects. Assigning precise thermodynamic loci would require a refined low-energy reference grid, bin-origin checks, and independent spectral-edge estimates.

\section{Reduced dynamics of coexisting pulses}
\label{app:reduced}
This appendix gives the conditional energy-sharing reduction summarized in Sec.~\ref{subsec:partition}.

\subsection{Adiabatic energy-sharing reduction}
\label{app:sharing}

This derivation uses intracavity energies. Multiplying every pulse energy by the fixed output-coupling factor leaves $\chi$ and the decay-rate ratio unchanged. For an unseeded deterministic field with periodic or vanishing boundary terms, Eq.~(\ref{eq:cgle}) gives
\begin{equation}
\frac{1}{2}\frac{dE}{dz}=-\sigma E-\alpha\!\int|\partial_ta|^2dt
+\kappa\!\int|a|^4dt-\kappa\zeta\!\int|a|^6dt.
\label{eq:energy_balance}
\end{equation}
The last three terms depend on shape as well as energy. Suppose, as an adiabatic approximation, that each well-separated pulse relaxes rapidly onto a single shape family, and that their sum, divided by $E_j$, becomes a differentiable function $G(E_j)$. Then $\dot E_j=2E_j[G(E_j)-\sigma]$, and stationarity requires
\begin{equation}
G(E_1)=\cdots=G(E_N)=\vartheta\left(\frac{\sum_j E_j}{E_{\rm cw}}-1\right).
\label{eq:partition}
\end{equation}
This closure excludes independent slow width, chirp, background, and interaction modes.

With $x$, $F(x)$, and $\chi(x)$ defined in Eq.~(\ref{eq:elasticity}), we can linearize the reduced equations about equal sharing. At equal sharing $E_j=E_c=F(x)$, differentiation along that branch gives $G'(E_c)=(N\vartheta/E_{\rm cw})(1-1/\chi)$. On a smooth branch with finite, nonzero $\chi$, linearization yields the following rates, defined as positive for decay. They are predictions of the stated closure, not eigenvalues computed from the full CGLE:
\begin{equation}
\begin{aligned}
\lambda_\parallel&=2E_c\left(\frac{N\vartheta}{E_{\rm cw}}-G'(E_c)\right),\\
\lambda_\perp&=-2E_cG'(E_c),\qquad
\frac{\lambda_\perp}{\lambda_\parallel}=1-\chi(x).
\end{aligned}
\label{eq:sharing_rates}
\end{equation}

\supplementarytitle{Supplementary Material: Energy-selective control of noise-assisted multipulsing by weak optical seeding in the dissipative-soliton-resonance regime}{Vladimir L. Kalashnikov, Alexander Rudenkov, Evgeni Sorokin, and Irina T. Sorokina}
\section{Seed timing and the seeded CW stage}
\label{supp:historical}

The timing routine draws an independent delay $\tau_n$ on every round trip, uniform over an interval $T_{\rm win}/q_\tau$ [rms $T_{\rm win}/(\sqrt{12}q_\tau)$] or Gaussian; $\sigma_\tau$ is thus a spread of delays, not a timing diffusion. For the envelope of main-text Eq.~(\ref{eq:seed0}), $S_0(t-\tau)-S_0(t)=2S_0(0)\sin[k(t-\tau/2)]\sin(k\tau/2)$ with $k=\pi/2T_s$, so that over the window
\[
\frac{|S_0(t-\tau_n)-S_0(t)|}{S_0(0)}\le\left(\frac{\pi T_{\rm win}}{2T_s}\right)^{2}\frac{2q_\tau+1}{8q_\tau^{2}}.
\]
For $T_s=0.8~\mu$s and $q_\tau=1$ the bound is $6.2\times10^{-7}$: timing jitter leaves such a seed unchanged. Delays comparable with $T_s$ would instead modulate the in-window seed amplitude by $\cos(\pi\tau_n/2T_s)$. The linewidth-derived option, $\sigma_\tau=\sigma_\phi/\omega_0=4.8$~fs for $\sigma_\phi=3.88$~rad, is below the 10-fs time cell; the linewidth acts through the phase [main-text Eqs.~(\ref{eq:phasejitter}) and (\ref{eq:phaserms})].

\begin{figure*}[t]
\centering
\includegraphics[width=0.82\textwidth]{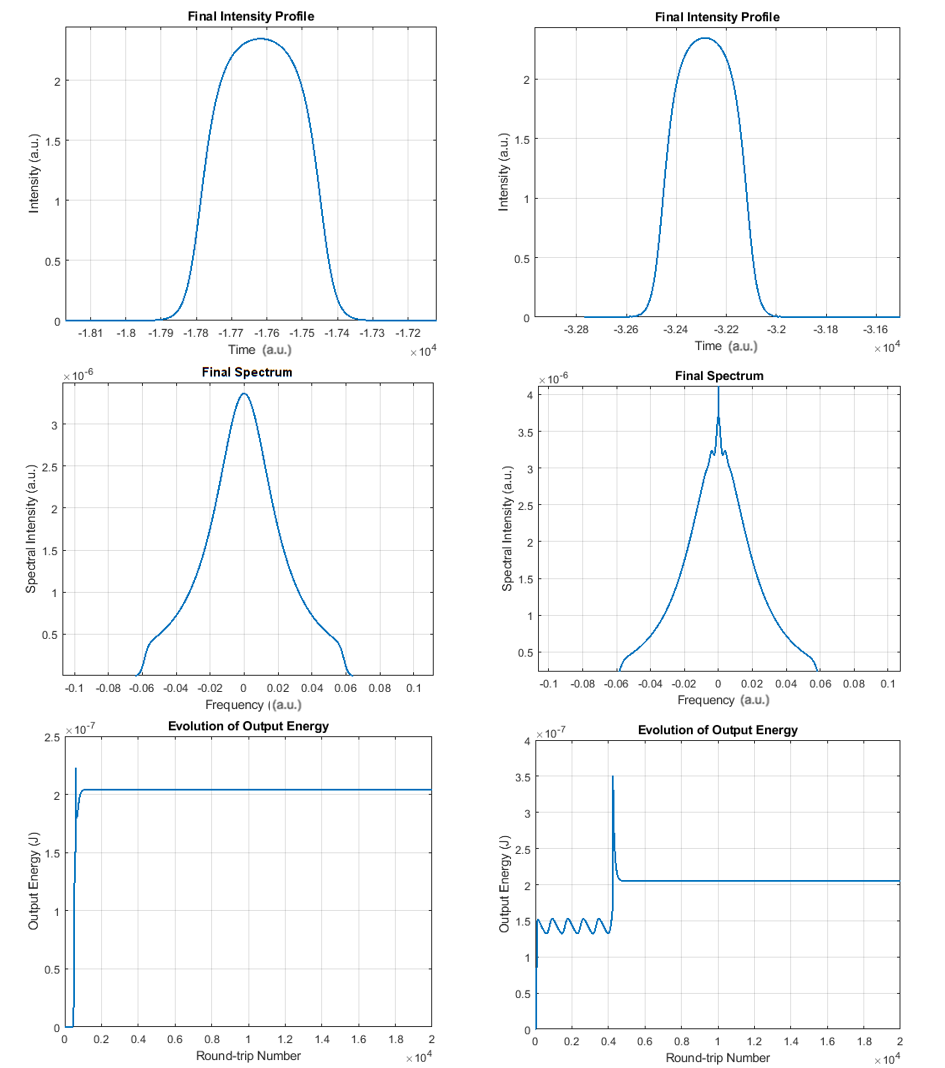}
\caption{Representative temporal profiles, spectra, and energy traces at $C=0.65$, $E'=15$, without injection (left) and with $A=10$~mW (right). Temporal, frequency, and intensity axes use the normalized simulation units. Energies are output energies ($\mathrm{OC}=0.14$).}
\label{fig:historical_profiles}
\end{figure*}

\begin{table}[t]
\caption{Driven-CW estimates for the low-energy fixed-phase conditions ($C=0.65$, $w=1$, 32 realizations each; $E_{\rm cw}/E'=65.5$~nJ) and for Fig.~\ref{fig:historical_profiles}. $\rho$: locking ratio; $\mu$: largest modulation growth rate ($10^{-3}$ per round trip); $10^4\mu$: e-foldings available in the run. Outcome after $10^4$ round trips; ``late'': pulse energies, not yet stationary.}
\label{tab:driven_cw}
\begin{ruledtabular}
\begin{tabular}{lcccl}
$A$, $E'$ & $\rho$ & $\mu$ & $10^4\mu$ & Outcome\\
\hline
1~mW, 5 & 0.62 & 1.24 & 12.4 & CW, oscillating\\
10~mW, 5 & 1.69 & $-3.31$ & --- & CW, stationary\\
10~mW, 7.5 & 1.02 & $\approx0$ & --- & CW, stationary\\
10~mW, 8.5 & 0.88 & 1.88 & 18.8 & CW, oscillating\\
10~mW, 9 & 0.80 & 1.94 & 19.4 & pulse, late\\
1~mW, 10 & 0.21 & 2.05 & 20.5 & pulse, late\\
10~mW, 10 & 0.68 & 2.59 & 25.9 & pulse, stationary\\
\hline
Fig.~\ref{fig:historical_profiles} & 0.36 & 4.12 & --- & pulse at 4200\\
\end{tabular}
\end{ruledtabular}
\end{table}

In Fig.~\ref{fig:historical_profiles}, the unseeded field forms a pulse after $\approx550$ round trips. With $A=10$~mW the field first becomes a uniform CW (within $\approx150$ round trips) whose energy oscillates between 0.132 and 0.153~$\mu$J with a period of 852--861 round trips; a pulse with the unseeded single-pulse energy forms only after $\approx4200$ round trips. For a uniform field $a=\sqrt{P}e^{i\theta}$, main-text Eqs.~(\ref{eq:cgle}) and (\ref{eq:saturation}) reduce to
\[
\begin{aligned}
\frac{dP}{dz}&=2P\bigl[-\sigma+\kappa P(1-\zeta P)+g_S\cos\Phi\bigr],\\
\frac{d\Phi}{dz}&=\gamma P-g_S\sin\Phi,
\end{aligned}
\]
where $\Phi$ is the seed phase relative to the field and $g_S=|S_0|/\sqrt{P}$. By Adler's equation \cite{sm-adler1946study}, the CW locks to the seed if $\rho\equiv|S_0|/(\gamma P^{3/2})\ge1$ and otherwise beats with the period $2\pi[(\gamma P)^2-g_S^2]^{-1/2}$. Fixing $E_{\rm cw}=0.98~\mu$J by the mean CW energy ($P=1.55$~kW, $\rho=0.36$) gives a period of 852 round trips and an energy range of 0.132--0.153~$\mu$J, as observed. The same $E_{\rm cw}/E'=65.5$~nJ reproduces the CW powers of the fixed-phase ensemble within 2\%.

A modulation $a=\sqrt{P}e^{i\theta}(1+p+iq)$ at frequency $\Omega$ obeys
\[
\begin{aligned}
\frac{dp}{dz}={}&(2g-g_S\cos\Phi-\alpha\Omega^2)\,p\\
&+(g_S\sin\Phi+\beta\Omega^2)\,q,\\
\frac{dq}{dz}={}&-(2\gamma P+g_S\sin\Phi+\beta\Omega^2)\,p\\
&-(g_S\cos\Phi+\alpha\Omega^2)\,q,
\end{aligned}
\]
with $g=\kappa P(1-2\zeta P)$. The growth rate is $2g$ without the seed and $g-g_S\cos\Phi$ for a locked CW. For an unlocked CW, $\mu$ is the Floquet exponent over one beat period, maximized over $\Omega$ ($\alpha$ and $\beta$ from main-text Table~\ref{tab:parameters}). Direct integration of main-text Eq.~(\ref{eq:cgle}) reproduces these rates within 1\%.

Spectral entropy separates CW-dominated endpoints ($\le0.24$~nats) from pulses ($\ge6.8$~nats) with no values between; classifying by it before counting corrects the 95 of 128 CW endpoints that the peak counter labeled multipulse and applies equally to repeated jitter runs. Table~\ref{tab:driven_cw} then shows three regularities. (i) $\rho$ separates the stationary CW endpoints ($\rho>1$) from the oscillating ones ($\rho<1$). (ii) The locked CW is stable at 10~mW and $E'=5$ and neutral at $E'=7.5$. (iii) Unlocked CWs grow: endpoints still CW at $10^4$ round trips had at most 18.8 e-foldings available and those with pulses at least 19.4, while the pulse of Fig.~\ref{fig:historical_profiles} formed after 17.4. All lie near $\ln(N_t^{1/2}|S_0|/\delta_0)=18.0$ (16.9 at 1~mW), the logarithm of the initial seed-to-noise amplitude ratio per Fourier mode [$\delta_0$: initial noise per sample at $w=1$, main-text Eq.~(\ref{eq:initial_noise_scan})]. Because every realization starts from the same seeded CW, each condition is uniform over its 32 runs.

The low-energy CW outcomes are thus seed-locked CWs or unlocked CWs that grow too slowly to form a pulse within the run, the counterpart of the competition between injection-locked emission and mode locking in quantum-dot lasers \cite{sm-rebrova2010stabilization,sm-habruseva2011dynamics}. Continued unlocked runs should form single pulses within $\approx1.5\times10^4$ round trips while $\Phi$ slips by $2\pi$ per beat; locked runs should stay CW with constant $\Phi$. Since $P=E_{\rm cw}/T_{\rm win}$, a longer window at fixed $E'$ raises $\rho\propto T_{\rm win}^{3/2}$ and lowers $\mu$, widening the locked range. In a physical cavity $\gamma P$ is negligible, and locking is limited by the seed--cavity detuning \cite{sm-wieczorek2005dynamical}.

\section{Rates and a modulated-seed test}
\label{supp:rates}
For two well-defined metastable classes with stationary transition rates,
\begin{equation}
\dot p_2=k_{12}(1-p_2)-k_{21}p_2,
\qquad
p_2^{\rm st}=\frac{k_{12}}{k_{12}+k_{21}}.
\label{eq:twostate}
\end{equation}
The fraction relaxes to $p_2^{\rm st}$ at $\lambda=k_{12}+k_{21}$. Continuing saved single- and multipulse endpoints under the same seed and noise measures both rates; equal late fractions from the two preparations validate Eq.~(\ref{eq:twostate}), and resolving pulse number and separation turns it into a rate matrix.

For weak noise the rates take the Kramers form
\begin{equation}
k_{ij}\simeq K_{ij}\exp[-\mathcal S_{ij}/D],
\end{equation}
with a noise convention and projected escape action specified consistently \cite{sm-kramers1940brownian,sm-RevModPhys.62.251,sm-grafke2019instanton}. Slopes of $\ln k_{ij}$ versus $1/D$ give the actions, and $\mathcal S_{21}-\mathcal S_{12}=D\ln(k_{12}K_{21}/k_{21}K_{12})$ measures the seed's energy selectivity as a barrier difference. Scanning the in-run kicks at a fixed initial field, with $D$ from main-text Eq.~(\ref{eq:qnoise_power}) for each class, separates activated escape from initial-condition effects.

Stochastic resonance \cite{sm-RevModPhys.70.223} is tested by modulating the seed, $A_n=A_0[1+\varepsilon\cos(\Omega_m n)]$. With $\eta_{ij}=\partial\ln k_{ij}/\partial\ln A$, Eq.~(\ref{eq:twostate}) gives
\[
\begin{aligned}
\delta p_2(n)={}&\varepsilon\,p_2^{\rm st}(1-p_2^{\rm st})(\eta_{12}-\eta_{21})\\
&\times\mathrm{Re}\,\frac{\lambda\,e^{i\Omega_m n}}{\lambda+i\Omega_m}.
\end{aligned}
\]
Because $\eta_{12}-\eta_{21}\propto1/D$ while $\lambda$ decreases exponentially in $1/D$, the amplitude peaks where $\lambda(D)\sim\Omega_m$, and the phase lag $\arctan(\Omega_m/\lambda)$ measures $\lambda$. At $T_{\rm cav}=0.8~\mu$s, $\Omega_m=10^{-3}$ per round trip is 200~Hz; modulated CW injection is established experimentally \cite{sm-bao2015switching,sm-zou2025resonantly}.

\section{Ensemble coherence}
\label{supp:coherence}
For independently sampled complex fields, the mutual-coherence kernel and participation number are \cite{sm-wolf1982partial,sm-starikov1982effective}
\begin{equation}
J(t_1,t_2)=\langle a_n(t_1)a_n^*(t_2)\rangle_n,
\qquad N_{\rm PR}=\frac{(\operatorname{Tr}J)^2}{\operatorname{Tr}(J^2)}.
\end{equation}
A deterministic field gives $N_{\rm PR}=1$ for any chirp or pulse count, so $N_{\rm PR}$ counts shot-to-shot field patterns, complementing the single-pulse spectral ratio $r$.

Within a pulse class, the decrease of $N_{\rm PR}$ from the raw to the time-aligned and then the frequency-aligned ensemble measures position and carrier wandering; the remainder measures shape and chirp variability (global phase cancels in $J$). With 32 realizations, $N_{\rm PR}\le32$; convergence is checked on random subsets of 8--32 realizations with bootstrap intervals.

\section{Coupling of seed timing and phase to a pulse}
\label{supp:projection}
Projecting the seed perturbation on an adjoint mode $\psi_j^\dagger$ of the linearized pulse \cite{sm-206583,sm-paschotta2004noise} gives the timing and phase weights
\begin{equation}
M_j^{(t)}=|\langle\psi_j^\dagger,S'_0\rangle|^2,\qquad
M_j^{(\phi)}=|\langle\psi_j^\dagger,iS_0\rangle|^2.
\label{eq:projection}
\end{equation}
Near the window center $S_0'\simeq-k^2tS_0(0)$, so $M_j^{(t)}/M_j^{(\phi)}\simeq k^4|\langle\psi_j^\dagger,t\rangle|^2/|\langle\psi_j^\dagger,1\rangle|^2$; the timing perturbation is at most $6\times10^{-7}$ of the seed (Sec.~\ref{supp:historical}), whereas a 3.9-rad phase step changes it by order unity, so phase fluctuations dominate the forcing. Pinning by timing control would appear as decay of an imposed pulse displacement, and locking as a fixed pulse phase relative to the seed.

Between pulses the seed is damped by $\sigma_b=\vartheta(E_1/E_{\rm cw}-1)$, 0.019 per round trip at $E'=15$. A fixed-phase seed builds a coherent background $|S_0|/\sigma_b$; a seed decorrelated every round trip ($e^{-\sigma_\phi^2/2}=5\times10^{-4}$ at 3~MHz) builds only $|S_0|/(2\sigma_b)^{1/2}$, $(2/\sigma_b)^{1/2}=7$--15 times less for $E'=5$--45. This parallels Ref.~\cite{sm-korobko2023birth}: an in-phase CW stabilizes one soliton, a phase-shifted CW triggers a second.

Langevin equations for energy, position, and phase, with damping rates from the linearized pulse at $E'/N$ and forcing weights from Eq.~(\ref{eq:projection}), form the reduced model; growth of the residual continuum marks background nucleation, the first step toward an additional pulse.

\section{Targeted tests of multipulse attraction}
\label{supp:attraction}
Pairs prepared at two separations persist with energies near $2E_1(E'/2)$, consistent with weakly interacting families. Four tests characterize these states.

(i)~\textit{Count.} Re-extract close pulses from the saved fields by resolved valleys and background contrast, varying the joining scale and smoothing; keep an ambiguous class; use the reference at $E'/N$ for each $N$; resolve drift between the saved snapshots (8000--$10^4$ round trips) by sampling every 10 round trips.

(ii)~\textit{Attraction.} Continue endpoints deterministically after small energy-asymmetry, relative-phase, and separation perturbations, removing only global symmetries: return to the same separation indicates an isolated bound state, to a nearby one a family.

(iii)~\textit{Energy sharing.} The main-text reduction predicts that an energy-conserving imbalance relaxes at $1-\chi(E'/2)=0.29$, 0.28, 0.25, and 0.22 times the rate of a symmetric perturbation at $E'=15$, 20, 30, and 40 ($C=0.65$); per-pulse energies resolve this slow transfer, and larger imbalances then probe the basin.

(iv)~\textit{Robustness and control.} Refine grid, step, and window at fixed physical noise covariance (window dependence of a CW: Sec.~\ref{supp:historical}); extend runs and parameters to separate families, instabilities, and transients; compare with an unseeded ensemble under the saved-field protocol.

\section{Experimental test}
\label{supp:experiment}
A seed is phase stable for $\Delta\nu_L\ll(2\pi T_{\rm cav})^{-1}=0.2$~MHz [main-text Eq.~(\ref{eq:phaserms})]; the 3-MHz case is the decorrelated limit. It stays locked to a cavity mode within $\pm g_S/(2\pi T_{\rm cav})$, $g_S=(P_{\rm inj}/P)^{1/2}$: about $\pm20$~kHz for 10~mW injected into the 1.2-W CW of $E_{\rm cw}=0.98~\mu$J at $T_{\rm cav}=0.8~\mu$s, which sets the required cavity-length stability. Repeated starts versus pump power, with the seed off, phase stable, and phase diffusing, should show (i)~an extended single-pulse range [main-text Fig.~\ref{fig:seeded_energy_scans}(a)], (ii)~alternating windows at stronger injection [Fig.~\ref{fig:seeded_energy_scans}(b)], (iii)~injection-locked CW competing with mode locking at low pump \cite{sm-rebrova2010stabilization}, (iv)~a locked or beating CW stage before pulse formation, and (v)~the seed line in the mode-locked spectrum (Fig.~\ref{fig:historical_profiles}). Modulating the seed at 20--200~Hz (rates of $10^{-4}$--$10^{-3}$ per round trip) implements the test of Sec.~\ref{supp:rates}.

\begin{supplementaryreferences}

\end{supplementaryreferences}

\end{document}